# Large Teams Have Developed Science and Technology; Small Teams Have Disrupted It

**Authors:** Lingfei Wu[1,2,3], Dashun Wang[4,5], James A. Evans[1,2,3]*

**Affiliations:**
[1] Department of Sociology, University of Chicago, 1126 E 59th St, Chicago, IL 60637.
[2] Knowledge Lab, University of Chicago, 5735 South Ellis Avenue, Chicago, IL 60637.
[3] Computation Institute, University of Chicago, 5735 South Ellis Avenue, Chicago, IL 60637.
[4] Kellogg School of Management, Northwestern University, 2001 Sheridan Rd, Evanston, IL 60208.
[5] Northwestern Institute on Complex Systems, 600 Foster Street, Evanston, IL 60208.
*Corresponding author. E-mail: jevans@uchicago.edu (J.A.E.)

**Abstract:** Teams dominate the production of high-impact science and technology. Analyzing teamwork from more than 50 million papers, patents, and software products, 1954-2014, we demonstrate across this period that larger teams *developed* recent, popular ideas, while small teams *disrupted* the system by drawing on older and less prevalent ideas. Attention to work from large teams came immediately, while advances by small teams succeeded further into the future. Differences between small and large teams magnify with impact—small teams have become known for disruptive work and large teams for developing work. Differences in topic and research design account for part of the relationship between team size and disruption, but most of the effect occurs within people, controlling for detailed subject and article type. Our findings suggest the importance of supporting both small and large teams for the sustainable vitality of science and technology.

**One Sentence Summary:** Data on more than 50 million teams in science and technology reveal systematic, fundamental differences between works produced by small and large teams: Across a wide variety of domains, small teams tend to disrupt science and technology with new ideas and opportunities, while large teams develop existing ones, suggesting the importance of both for continuing advance through discovery and invention.

One of the most universal shifts in science and technology today is the flourishing of large teams in all areas of science, scholarship and invention as solitary researchers and small teams diminish (*1–4*). Increases in team size are attributed to the specialization of scientific activities (*4*), technological advances that lower communication costs (*5*, *6*), and the emergence of large, highly connected knowledge communities (*1*, *5*). The flourishing of large teams in an environment historically populated with small teams and solo investigators raises an important question: How do large and small teams differ in the character of the science and technology they produce? Here we test the hypothesis that small teams are more likely to *disrupt* science and technology with new problems and opportunities, while large teams tend to *develop* them with solutions and refinements, a dichotomy with many titles (*7–11*). To this end, we analyze teamwork and citation patterns from more than 50 million research articles, patents, and software to systematically explore the nature of collaborative activity across diverse domains.

Past research has shown that article and patent citation counts post a slightly positive, high variance relationship with team size (*2*, *12*). Citation counts alone, however, cannot capture distinct contributions, which we illustrate with three examples shown in Fig. 1A-C. We visualize three well-known articles with similar impact, but very different contributions (*13–15*). The Bak *et al* article on self-organized criticality (commonly referred to as the BTW model) (*13*) received a similar number of citations to the Davis *et al* article on Bose-Einstein condensation (*14*), but most research subsequent to Bak *et al* only cited the BTW model, without mentioning its references (green links in Fig. 1A). In contrast, Davis *et al*, for which Wolfgang Ketterle was awarded the 2001 Nobel Prize in Physics, is almost always co-cited with its antecedents (brown links in Fig. 1B). The difference between the two papers is not reflected in citation counts, but in whether they disrupt or develop existing scientific ideas—suggest or solve a scientific problem. The BTW model launched new streams of research, while the experimental realization of Bose-Einstein condensation elaborated formerly posed possibilities.

To systematically evaluate the role that scientific and technical work plays in unfolding advance, we collected large-scale datasets from three related but distinct domains (SOM Materials and Methods): (1) Web of Science database containing more than 43M articles published between 1900 and 2014 and 615M citations among them; (2) Patents granted by the U.S. Patent & Trademark Office (USPTO) from 2002 to 2014 and citations added by patent applicants; (3) Software projects on GitHub, a popular web platform that allows users to collaborate on the same code repository and also "cite" other repositories by forking and building on their code. We analyzed core memberships for each repository and forking patterns among them.

For each dataset, we assess citation structure in three distinct ways. First, we measure the degree to which a work disrupts the field of science or technology by introducing a new idea that eclipses attention to the prior art it draws upon (*16*). The measure varies between -1 and 1, corresponding to work that develops or disrupts, respectively (Fig. 1B-D). Second, we measure delay in attention to science and technology using the "sleeping beauty index" (*17*, *18*), which captures a delayed burst of attention by calculating convexity in a work's citation distribution over time. The index is highest when a paper is never cited for some period before receiving its maximum, corresponding to belated appreciation, 0 if cited linearly in the years following publication, and negative if citations chart a concave function with time tracing early fame, diminishing thereafter. Finally, we examine impact by the number of citations the work received. If large team contributions develop existing ideas, scientists and engineers alongside them constitute a ready market for these developments. We predict that small team work will be more disruptive, receive citations after a longer delay, and collect less citations overall due to the rapid decay of collective attention (*19*, *20*).

The three settings we studied differ clearly in their scope, domain, and typical time scales, but we consistently find that outputs by teams of different size have played distinctive roles in advance. Large teams have tended to produce articles, patents, and software that garner modestly higher impact, but the disruption of these products dramatically and monotonically declines with each additional team member (Fig. 2A-C). As teams grow, the likelihood that they eclipse the work on which they build vanishes. Specifically, as teams enlarge from one to fifty team members, their papers, patents and products drop in disruption by 70%, 30% and 50%, respectively. In every case, this highlights a dramatic transition from disruption to development as disruption curves drop below the dashed line marking the zero point. These results support the hypothesis that large teams may be better designed or incentivized to develop current science and technology, while small teams disrupt it with new problems and opportunities.

We uncover the same conclusion when we focus on only the most disruptive and impactful works (Fig. 2D—F). As shown in Fig. 2D, solo authors are just as likely to produce a hit paper (top 5% citations) as teams with five members, but their articles are 72% more likely to be highly disruptive (top 5% disruption). In contrast, ten-person teams are 50% more likely to score a hit paper, yet these contributions tend to develop existing ideas already prominent in the system, as reflected in the very low likelihood they are among the most disruptive. Repeating the same analyses for patents (Fig. 2E) and software development (Fig. 2F), we find that disruption and impact universally diverge as team size grows.

Disruption differences between small and large teams magnify with impact (Fig. 3A). Small teams producing high-impact papers are most disruptive, and large teams producing high-impact papers most developmental. As article impact increases, the negative slope of disruption as a function of team size steepens sharply. Even within the pool of high impact articles and patents (Fig. 3A, top 5% citations), which are statistically more likely produced by large teams (Fig. 2D), small teams have disrupted the current system with substantially more new ideas. Beyond impact level, we further split papers by detailed scientific field (Fig. 3B and S9-S17) and time period (Fig. S4), finding that these patterns hold remarkably stable for all eras and 90% of the disciplines. The only consistent exceptions were observed for disciplines (e.g., Engineering and Computer and Information Technology) where conference proceedings rather than journal articles are the publishing norm (our WOS data only indexes journal articles).

Part of the difference between small and large teams is surely due to differences in the topic, research design and resources required for distinctive work each performs, rather than a causal factor of team size itself. Review articles are typically crafted by single authors or small teams, but massive experiments demand the coordination and lobbying power of an entire community. We controlled for author differences by comparing the same author's articles against themselves, varying only team size (Fig. 3C), and we modeled this relationship accounting for a hundred variables that detail the coordinates of each article's title and abstract in the high-dimensional space of published science (see SOM; *21*). These comparisons and models reveal that approximately one third of the team size effect we find can only be observed across different scientists presumably doing different kinds of science. Moreover, different kinds of science strongly influence the degree to which articles disrupt or develop science, increasing our model fit by an order of magnitude. Nevertheless, we continue to observe nearly two thirds of the effect shown in Fig. 2 when we compare scientists with themselves, varying team size and content. We also find the same patterns when we exclude review articles (Figure S7), and when we consider review articles alone, with reviews written by several authors substantially less disruptive than those written by few.

The considerable difference in disruption between large and small teams raises questions regarding how these teams differ in searching the past to formulate their next paper, patent or product. When we dissect search behavior, we find that large and small teams engage in strikingly different practices that lead to divergent contributions in disruption and impact. Specifically, we measure *search depth* as the average relative age of references cited (*32*) and *search popularity* as the median citations to a focal work's references. We examine these search strategies and consequences across fields, time periods, and impact levels in science, technology and software. We find that solos and small teams are much more likely to build on older, less popular ideas (Fig. 2G-L).

Larger teams, with more people spanning more dispersed areas, cannot be less aware of older, less popular work than small teams, but they have been systematically less likely to build on it. Indeed, larger teams have been much more likely to target recent, high-impact work as the primary source of their ideas, and this tendency increases monotonically with team size. It follows that large teams have received more of their citations rapidly, as their work is immediately relevant to more contemporaries whose ideas they develop. Conversely, smaller teams experience a much longer citation delay, with an average Sleeping Beauty Index for solo and two-person research teams four times that of ten-person teams (Fig. 3D). Our findings also reveal a ripple effect, whereby successful small team research becomes the basis for later large team success (Fig. S29).

An often claimed advantage of large teams is their ability to link divergent fields (*22*, *23*). We find that this effect grows as a convex function of team size. The effect of broader teams on fusing surprising combinations from diverse journals saturates between eight and ten team members and then reverses with greater team size, dropping below solo authors and smaller teams (Fig. S4). These results suggest that combinations of distant ideas are benefited by broad teams, but that they are more likely to enter published research when they occur *within* a few team members' individual experiences than *across* the experiences of many team members.

In summary, we report a universal, previously undocumented pattern that systematically differentiates the contributions of small and large teams in the creation of scientific papers, technology patents and software products. Small teams have disrupted science and technology by exploring and amplifying promising ideas from older and less popular work. Large teams have developed recent successes, solving acknowledged problems and refining common designs. Part of these differences result from differences in the substance of the science that small versus large teams tackle, and part appear to result from the structure of team size itself. Certain types of research require the resources of large teams, but large teams demand an ongoing stream of funding and success to "pay the bills" (*24*) and they may be more sensitive to the risk over the loss of reputation and support from failure (*25*). Our findings are consistent with field research on teams in other domains, which demonstrate that small groups with more to gain and less to lose, tend to undertake new, untested opportunities, with potential for high growth and failure (*26*, *27*). Our findings also accord with experimental and observational research on groups that demonstrates how individuals in large groups think and act differently (*28–32*).

Both small and large teams are essential to a flourishing ecology of science and technology. The increasing dominance of large teams, a flurry of scholarship on their perceived benefits (*33–42*), combined with our findings call for new investigation into the vital role played by individuals and small groups in advancing science and technology. Direct sponsorship of selected small group research may not be enough to preserve their benefits. Analyzing articles published from 2008 to 2012 that acknowledged financial support from several top government agencies around the world, we find that the small teams receiving funds are indistinguishable from large teams in their tendency to develop rather than disrupt their fields (Fig. S30). This could result from a conservative review process, proposals designed to anticipate such a process, or a planning effect whereby small teams lock themselves into big team inertia by remaining accountable to a funded proposal. Regardless of the dominant driver, these results paint a unified portrait of bold, broke, solo investigators and small teams who disrupt science and technology by generating new directions based on deeper and wider information search. We recommend that government, industry and nonprofit funders of science and technology support the critical role small teams play in expanding the frontiers of knowledge, even as large teams rapidly develop them.

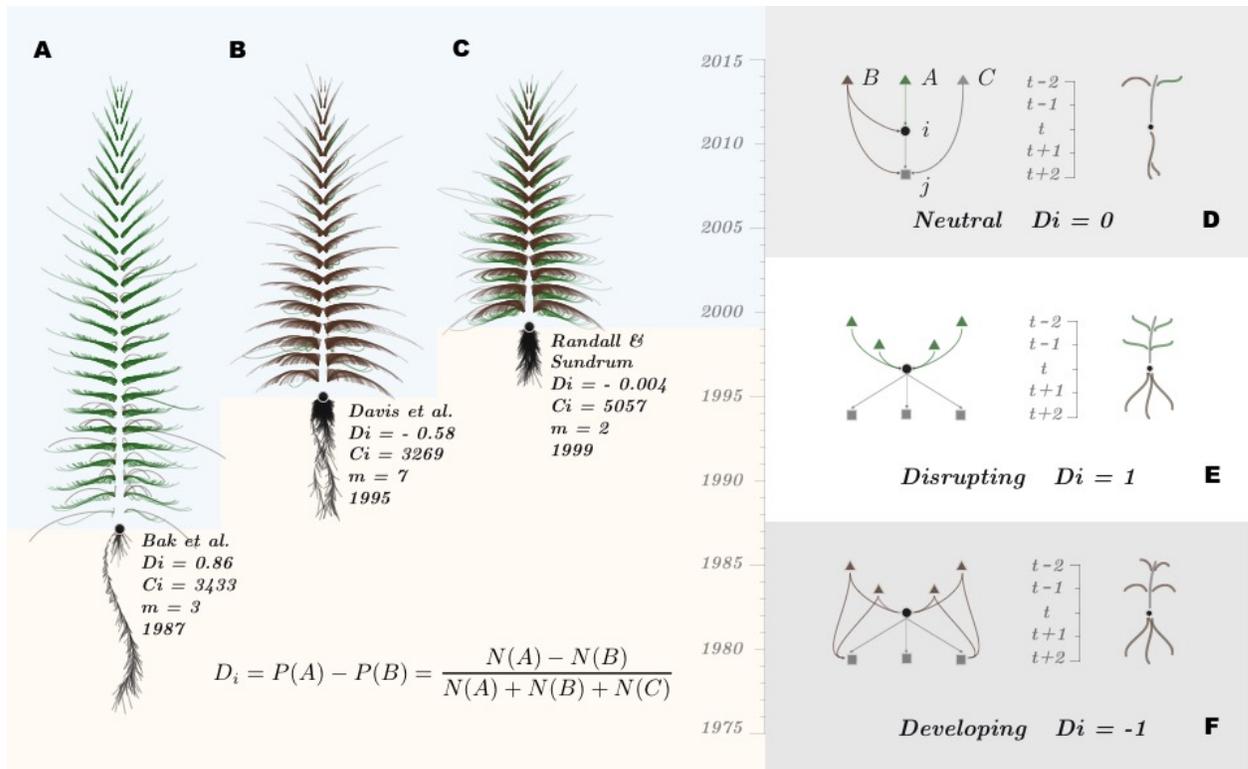

**Fig. 1. Disruptive and developmental papers.** (**A - C**) Three articles of the same impact scale $C_i$ represented as citation trees illustrate how disruption measure $D_i$ distinguishes different contributions to science and technology. "Self-organized criticality: An explanation of the 1/f noise" by Bak et al. (A), "Bose-Einstein Condensation in a Gas of Sodium Atoms" by Davis et al. (B), and "A Large Mass Hierarchy from a Small Extra Dimension" by Randall and Sundrum (C). Each draws on past work and passes ideas onto future work: "roots" in yellow zone are references, with depth scaled to publication date; "branches" in blue zone are citing articles, with height and length scaled to publication date and impact, respectively. Branches curve downward if citing articles also cite the focal paper's references, and upward if they ignore them. (**D**) Simplified illustration of disruption: Citation network comprising focal paper $i$ (black circle), reference $j$ (gray rectangle), and three subsequent works (triangles). A cites only focal work $i$; B cites both $i$ and prior, referenced work $j$, and C cites only $j$. The disruption of focal paper $i$ is defined by $D_i = P(A) - P(B)$ or $D_i = (1-1)/3 = 0$, suggesting that the work balances disruption and development by eclipsing prior work (to A) and amplifying it (to B). (**E**) When a work's novelty completely overshadows prior work by receiving all subsequent attention itself, then $D_i = 1$. (**F**) When a work is always cited alongside its inspirations, it primarily broadcasts the importance of prior work, hence $D_i = -1$.

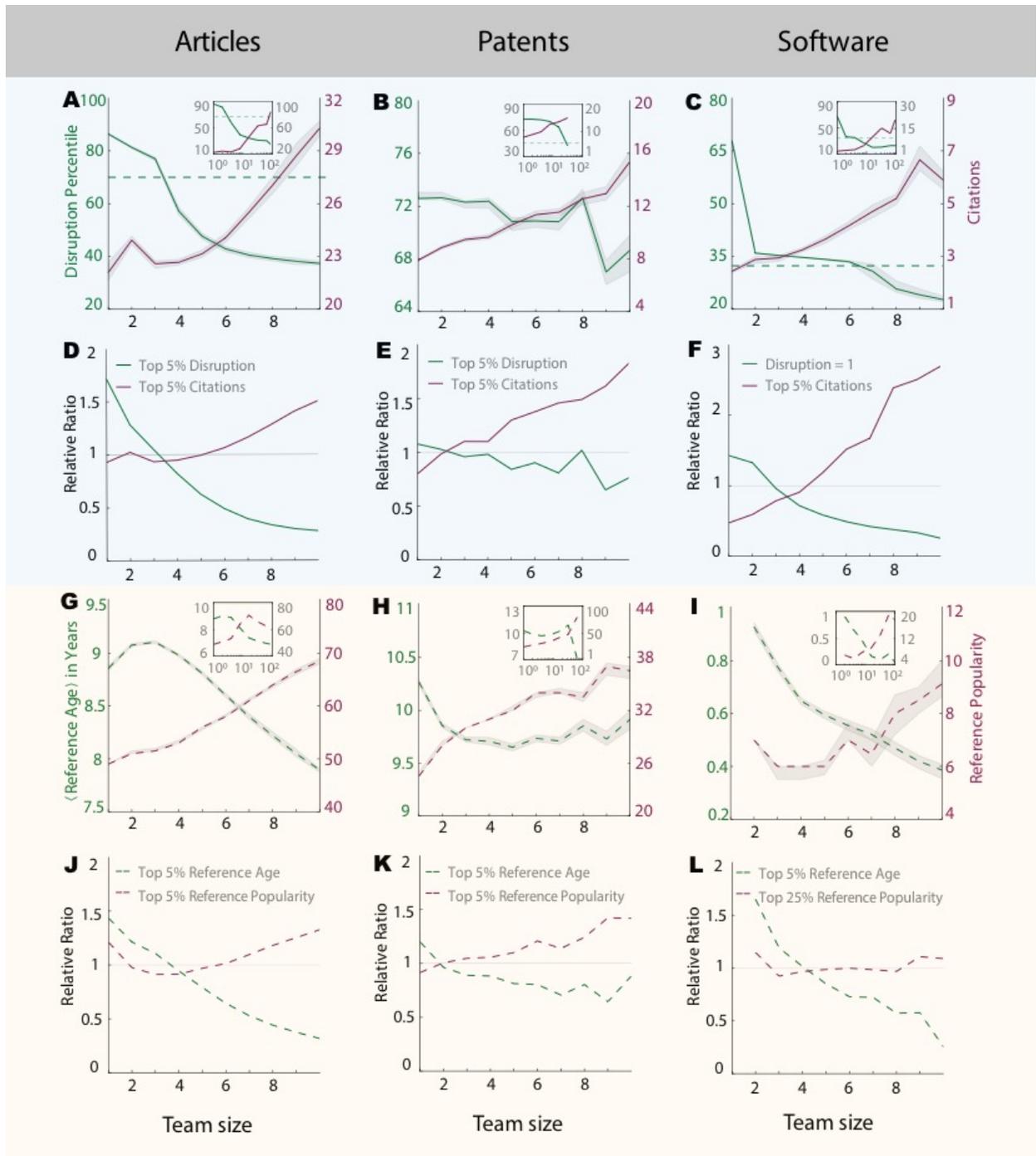

**Figure 2. Small teams disrupt, big teams develop.** (**A - C**) For research articles, patents, and software, average citations (red curves indexed by right y-axis) increase with team size, whereas disruption percentile (green curves indexed by left y-axis) decreases with it. For all data sets, we present work with one or more citations. Bootstrapped 95% confidence-intervals are shown as gray zones. Insets reveal that observed relationships hold for two orders of magnitude of team size. Green dotted lines show where $D_i = 0$, the transition from development to disruption. (**D - F**) Same as (A - C) but extreme cases

rather than average behavior. The probability of observing papers, patents and products of highest impact increases with team size, while probability of observing the most disruptive decreases with it. Relative ratio is the ratio of empirical percentages to those expected if teams were equally distributed in output qualities. In Software, 69% of the codebases have disruption values that equal 1, therefore we use this maximum value instead of the top 5%. (**G - I**) Influence on the future (A-C) relates to how teams search the past. Median popularity of references (in number of citations) increases with team size, while average age of references decreases with it. (**J - L**) Same as (G - I) but extreme cases. Software has very few high-citation codebases, and so we use top 25% rather than top 5% reference popularity.

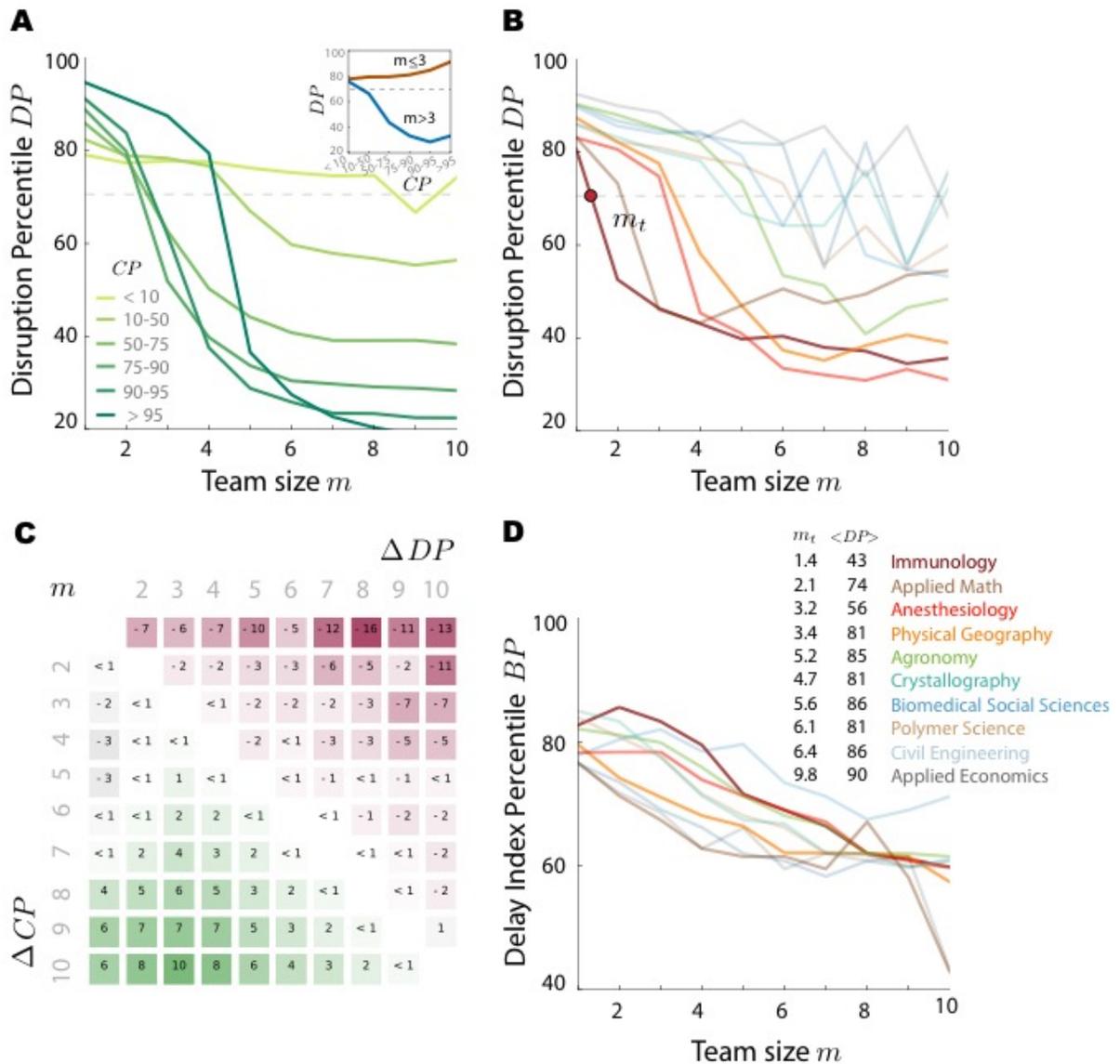

**Figure 3. Small teams disrupt across impact levels and fields.** (**A**) Disruption percentile *DP* decreases with team size *m* across impact levels. Curves are colored by impact percentiles (in number of citations). *DP* decreases faster for higher impact articles (darker green curves). The dotted gray line shows $D_i = 0$; $DP = 70$. The inset reveals how impactful small teams are distinguished by disrupting science and technology, while impactful large teams are distinguished by developing it. (**B**) Disruption decreases with team size across 10 subfields (see Fig. S19-27 for a systematic investigation of team size and disruption across 12 fields, 218 subfields, and 10,907 journals). For each field we can find a threshold of team size $m_t$ (red circle) at which point adding more team members makes the collaborative work transition from disruptive to developmental. (**C**) Shifts in the Citation and Disruptive percentile between articles published by the same author participating in groups of different size. (**D**) Our Delay or Sleeping Beauty Index

percentile, *BP* (*17*) measures the temporal delay of a work's biggest citation burst since publication and decreases dramatically with team size across fields. The legend shows field color scheme, ranked values of $m_t$, along with average *DP* and *BP*.

**Materials and Methods**

Web of Science data
Our Web of Science (WOS) dataset contains 43,661,387 papers (distributing across 15,188 journals) and 615,697,434 citations spanning 1900 - 2014. 76% papers have more than one author. The data before 1950 are sparse, and so results presented in the main text focus on papers published 1954 - 2014. The data from this period contain 42,045,077 papers, distributed across 15,070 journals. Among these articles, 66% are cited at least once, generating 611,483,153 citations in total. 78% of these papers have more than one author. To calculate disruption and other network measures (Fig. 2), we constructed a directed network with papers as nodes and citations as links.

U.S. Patent data
The U.S. Patent & Trademark Office patent data set (USPTO) contains 4,910,816 patents and 64,694,807 citations between 1976 – 2014, the portion of the dataset with curated, digital patent application data. 60% of these patents were authored by more than one inventor. The citation links added by inventors and patent examiners represent very different dynamics (see Figure S8), with examiner citations not reflecting technology on which a proposed invention built, but rather technologies with which it competes. As such, we focus only on applicant citations, which are marked in the data set after 2001 and represent 53% of total citations. From 2002 - 2014 we have 2,548,038 patents in total, linked by 44,798,680 inventor citations. To calculate disruption and other measures (Fig. 2), we constructed a directed network that contained patents as nodes and applicant citations as links. This resulting network contains 1,288,181 nodes (patents) generated in or after 2002, creating 23,854,363 links. 68% of these patents are contributed by more than one inventor.

GitHub data
The GitHub data contains 15,984,275 code bases (repositories) contributed by 2,348,085 programmers in GitHub between 2011 and 2014. To calculate disruption and other measures we construct a citation network of repositories. We add a citation link from repository A to B if a "core member" of A, user *i*, copied and saved ("forked") the code from B during the time period in which *i* was contributing to A (which lasts from *i*'s first to last edit of A). For each repository, we identify its core members as those who contributed more edits, or "pushes", than the average value of all contributors to a repository. The constructed network used to calculate disruption and other network measures (Fig. 2) contains 26,900 nodes (repositories) and 108,640 links.

Google Scholar data
Google Scholar data include 2,040,419 papers contributed by 88,863 scientists and scholars, collected by crawling Google Scholar (GS) profiles and then matching listed papers to articles in the Web of Science. The most productive scholar in our data contributed to 924 articles while many scholars (>10,000) contributed to only a single article. We use the Google Scholar data to validate our name disambiguation algorithm, which in turn allowed us to investigate the shift in impact and disruption that occurs *within* the same scientists and scholars involved in both smaller and larger teams (Fig. 3C).

Limitations of Datasets
We acknowledge that the three large-scale datasets being used in our study—Web of Science, U.S. patents and github software repositories—may contain biases. Some of these are known, which allow us to account for them in our analyses. For example, Computer Science and some engineering disciplines rely more intensively on conference proceedings to communicate key results, so journals are less relevant for these communities. In other cases, the biases remain "unknown unknowns". In the software case, for example, it is unknown whether free and open source software projects are uniformly represented on github, although an increasing number are indeed hosted in that environment. As with other large-scale, data-driven studies, these biases limit the scope and generalizability of our claimed findings. In the end, all we can assert within confidence is that these patterns hold across the millions of project that represented within the Web of Science, the U.S. patent database, and github public software repositories.

**Supplementary Text**

Background
Team science advocates have claimed that a shift to larger teams in science and technology fulfills an essential function: most problems that humankind faces in modern society are complex and require interdisciplinary teams to solve (*1–4*). As the knowledge frontier expands exponentially (*5*, *6*), the promise of teams—especially large teams—lies in their ability to reach across and more effectively combine knowledge to produce tomorrow's breakthroughs (*7*, *8*). While much has been discussed about the professional and career benefits of team science (*9*), it remains unclear whether being embedded in a large team is always a good strategy for knowledge discovery and technological invention (*9*, *10*). Experimental and observational research on groups reveals that under certain conditions individuals in large groups think and act differently—they generate fewer ideas (*11*, *12*), recall less learned information (*13*), reject external perspectives more often (*14*), and tend to neutralize each other's viewpoints (*15*). This led us to explore the relative differences in the contributions of small and large teams, their search processes, and how their work impacts the unfolding frontier of advance.

Journals, subfields, and fields
The articles we analyzed are published across 15,146 journals, which belong to 258 subfields according to the subject category labels for journals in the WOS dataset. We further group these subfields into 14 major fields comprising Medicine, Biology, Physical sciences, Engineering, Environmental and earth sciences, Chemistry, Social sciences, Agriculture, Business and management, Computer and information technology, Mathematics, Multidisciplinary Sciences (e.g., *Science*, *Nature*, *PNAS*), Humanities, and Law. In Figures S9-17 we plot the average disruption percentile against team size for each journal across 12 fields (results for Humanities and Law are not shown due to lack of data), 218 subfields and 10,907 journals (only journals with more than three data points are shown). We use ordinary least quare (OLS) regression to fit the relation between team size and disruption percentile. We find that among all studied journals, 86% post negative regression coefficients. If we only consider journals that publish a substantial number of articles or those for which the regression coefficient is significant, this fraction is higher: 91% journals of more than 3,000 articles show negative relationship between team size and disruption percentile, and 88% journals give significant negative regression coefficients.

Identifying articles sharing the same author(s)
For each article $i$ co-cited alongside other articles within the reference list of a subsequent article, we select the most relevant article $j$ co-cited with $i$ the maximum number of times. If $i$ and $j$ share at least one author name (the full name in WOS), we assume that it is the same scholar. From this process we obtain 3.9 million article pairs. For the 0.2 million pairs also included in the Google Scholar data, we test the accuracy of our name disambiguation approach. The rate that these pairs come from the same, self-identifying Google Scholar profile author is 92.5%. These relevant article pairs allow us to systematically compare articles contributed by different teams sharing the same author(s) across the entire WOS system (Fig. 3C).

Random effects versus fixed effects models of team size on disruption
We use Google Scholar data to investigate the impact of team size on disruption by comparing the effect of team size on disruption between vs. within individual scholars (see Fig. S18 for examples). To do this, we compare estimation of the influence of team size on disruption in random effects and author-fixed effects models. If different kinds of scientists and research require different sized teams, then we should see an influence on disruption from team size in the random effects model, but not the author fixed effects model. In contrast, if team size alone affects scientists and the science they perform, then variation in disruption by team size should post the same effect in both random effects and author fixed effects models.

More specifically, we construct the random effects model $D_{ir} = \alpha_1 + \beta_1 m_{ir} + \varepsilon_{ir}$, in which $D_{ir}$ represents disruption of the paper authored by scientist $i$ in the $r$th smallest team on which he or she collaborates and $m_{ir}$ is the size of that team—the disruption of papers from teams of the same size are averaged. Parameters $\alpha_1$ and $\beta_1$ are the intercept and team size coefficients, respectively. We similarly construct the author fixed effects model $D_{ir} - \overline{D_i} = \alpha_2 + \beta_2 ( m_{ir} - \overline{m_i} ) + \varepsilon_{ir}$ to control for differences in disruption by team size within individual scientists (*40*). We estimate the parameters of these two models as $\beta_1 = -0.00027$ (p-value < 0.001) and $\beta_2 = -0.00018$ (p-value < 0.001), suggesting that two-thirds of the total variation in disruption can also be observed within scientists. Therefore, for the same scientist, merely by shifting from small to large teams, the disruption metric of his/her work manifests a statistically significant difference. Sizes of these regression coefficients are small because the values of $D_i$ cluster around zero (Fig. S2D-E). Between these two models, the Durbin-Wu-Hausman (*41*) test favors the fixed model. In summary, these models suggest that approximately one third of the difference in disruption by team size may be accounted for by different kinds of scientists doing different kinds of science. Nevertheless, the substantial remaining difference in disruption is observed within scientists as they move between teams and projects of smaller and larger size.

Controlling for topics in science

We also compare the author fixed effects model against the random effects model while controlling for article topics. We concatenated titles and abstracts for 100,000 articles published by scholars in the Google Scholar dataset, and used them as the training corpus to train a shallow neural network to converts documents into vectors (Doc2vec) (21). To insure that these articles cover various topics, they were selected randomly across the 11,370 journals contributed by those scholars, weighted by the frequencies of articles from the journals in the Google Scholar dataset. We used the Gensim Python library to train the vector space with model parameters as follows: size=100 (the vector length), min_count=2 (the minimum frequency of words used in the training), iter=20 (the number of iterations over training corpus) (21). After training, we measured the similarity between documents in the training set by calculating the cosine similarity between their estimated vectors. We find that greater than 96% of the inferred documents are found to be most similar to themselves, which suggests that the trained Doc2vec model is working in a usefully consistent manner. To provide face validity for our model, we randomly select three documents and provide documents that register as most and least similar (Table S1-3).

Using the trained model, we infer the vectors for each of the 2,040,419 articles in the Google Scholar dataset, and use their position in each of the 100 dimensions as (100) control varibles, included in both the random and author fixed effects model, as described in the last section. We estimate the parameters of these two models as $\beta_3$ = -0.00022 (p-value < 0.001) and $\beta_4$ = -0.00017 (p-value < 0.001). Comparing $\beta_1$ and $\beta_4$ we conclude that more than 60 percent of the total variation in disruption can be observed within scientists publishing on the same broad topic. We recognize that this control may still not capture *detailed* differences between projects on which an author researches and publishes.

Summary Figures and Tables

Figures S1-S30 and Table S1-3 detail properties of the papers, patents and software products data in relation to team size, disruption, impact, delay, and the depth, breadth and popularity of search behavior.

Figure S1-3 discuss the calculation of disruption and compare it against alternative measures. Specifically, Figure S1 compares examiner and applicant citations in U.S. patent data. Figure S2 reveals the detailed distributions of impact and disruption with team size in WOS articles. Figure S3 contrasts alternative disruption measures in terms of self-consistency and relationship with delayed impact using WOS data.

Figures S4-20 control for a variety of variables in three datasets to show that the observed negative relationship between team size and disruption is robust. We control for six variables in our analysis of WOS articles, including time period, time window, article type (review vs non-review), journal and subfield, author, and topic. We also control for two variables associated with U.S. Patents, including patent class and recipient; and two variables linked to GitHub software, including programming language and code base size. Specifically, Figure S4 traces the temporal evolution of disruption across different time periods in WOS articles. Figures S5 and S6 show how decreases in disruption $D$ and increases in impact $C$ with team size $m$ are robust to increases in the time-window of observation for WOS articles published in a randomly selected year (1970) and for all years. Figure S7 compares the dynamics between review and non-review articles. Figure S8 demonstrates the weighted average technique to smooth data, which is used in Figure S9-17, in which the dependency of disruption on team size is systematically investigated at the journal level. Figure S18 illustrates our design strategy used to build models that regress disruption on team size, controlling for author and topic. Figure S19 details the team size influences from random and author fixed effects models estimated on WOS articles. Tables S1-3 give examples of similar and dissimilar articles suggested by our Doc2vec model. Figure S20 illustrates how the decrease of disruption with the increase of team size is universal across classes and recipients in U.S. patents. Figure S21 illustrate how the decrease of disruption with increase of team size appears universal across programming languages and code base sizes for GitHub software projects.

Figures S22-23 explore citation dynamics related with disruption. Specifically, Figure S22 details how the expected value of reference age $t$ decreases with team size $m$ while the expected value of reference popularity $k$ increases with it in WOS articles. Figure S23 graphs the shift in combinatorial novelty with team size in WOS articles.

Figures S24-29 analyze the impact of disruption on long-term citations and explore possible consequences from the decline in small teams. Specifically, Figure S24 compares citation dynamics across team sizes and levels of disruption in WOS articles. Figure S25 reveals a temporal risk-reward trade-off in number of citations for WOS articles as a function of different levels of disruption. Figures S26 and S27 show how the negative correlation between disruption $D(t)$ and impact $C(t)$ in the short term ($t \leq 10$ years) turns positive in the long term ($t > 10$ years) for WOS articles published in a randomly selected year (1970) and for all years. Figure S28 illustrates the decline of small teams over time in WOS articles and U.S. patents. Figure S29 shows a ripple effect, whereby successful small team research becomes the basis of later, large team success. Figure S30 reveals how small team research funded by several national and regional funding agencies is less disruptive than larger team research funded by the same agencies, and substantially less disruptive than unfunded research published in the same journals and time periods.

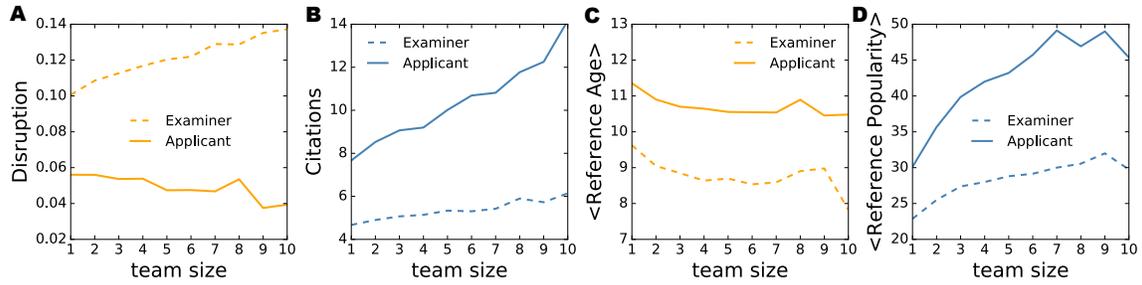

**Figure S1. Comparison of examiner and applicant citations in U.S. patent data.** Since 2002, citations required by examiners are noted in the USPTO dataset and allow differentiation from citations initiated by applicants. Here we construct two distinct networks using these two different citation types created 2002-2014. The applicant network has 3,709,872 nodes and 23,854,363 edges; 35% (1,288,181) of the cited patents were granted after 2001. The examiner network has 4,689,950 nodes and 12,276,780 edges; 46% (2,173,256) of the cited patents are generated after 2001. In this figure, we show results from a single year, i.e., 2009, to compare the dynamics between the two kinds of citations. (A) Disruption decreases with team size in the applicant network but increases with it in examiner network. (B) Number of citation increases with team size in both networks. (C) Average reference age decreases with team size in both networks. (D) Average popularity of references increases with team size in both networks.

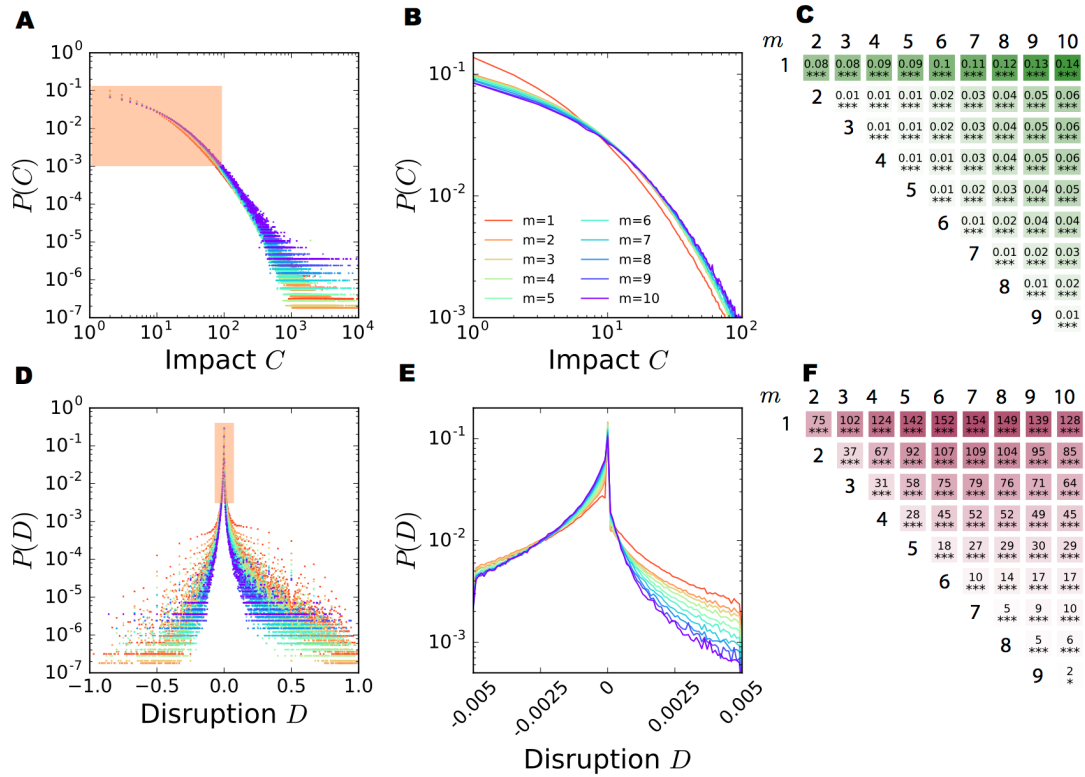

**Figure S2. The distribution of impact and disruption.** The distribution of impact (in citations) (A) and disruption (D) of WOS articles change with team size $m$ (color legend inset in B). (B) and (E) represent zoomed-in versions of the orange areas in (A) and (D), respectively. All figures clearly demonstrate how small teams over-sample less impactful and more disruptive work. In (C) we test difference in the distributions of impact between each pair of team sizes from one to ten using the two-sample Kolmogorov - Smirnov (KS) test. KS statistics are given in green cells, whose darkness is proportional to the numbers in the cells. The stars under the numbers indicate p-values: * $p \leq 0.05$, ** $p \leq 0.01$, and *** $p \leq 0.001$. In (D) we test the difference on distributions of disruption between team sizes using two-sample t-tests. Numbers in cells show t-statistics and the underlying stars indicate p-values. All distributions in C and F significantly differ from one another. (Comparing disruption distributions with the KS test reveals the same patterns of difference)

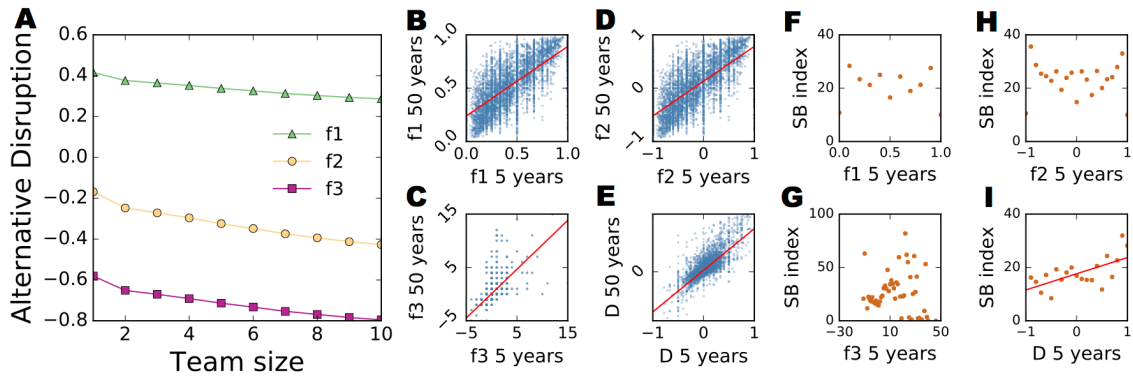

**Figure S3. Alternative disruption measures.** As shown in Fig. 1, disruption is defined as $D = P(A) - P(B) = (N(A) - N(B))/(N(A) + N(B) + N(C))$, in which A, B, C are three types of following papers and $N(A)$, $N(B)$, and $N(C)$ are fractions of these papers, respectively. (A) Three alternative measures, $f1 = N(A)/(N(A) + N(B))$, $f2 = (N(A)-N(B))/(N(A) + N(B))$, $f3=(N(A)-N(B))/N(C)$, are defined as alternative versions of disruption. All three variables decrease with team size, suggesting that changing the definition of disruption would not alter our findings. (B)~(E) We investigate the temporal robustness of $D$ and its alternatives by calculating the Pearson correlation between 5-year and 50-year values. The Pearson correlation is 0.80 for $f1$, 0.80 for $f2$, 0.33 for $f3$, and 0.78 for $D$. All correlation coefficients are significant. (F)~(I) We investigate the predictive power for delayed citation bursts (measured by the Sleeping Beauty Index) using 5-year $D$ and its alternatives. Only $D$ is significantly correlated with Sleeping Beauty Index. The Pearson correlation coefficient is 0.64, with a P-value 0.003. To summarize, while several alternative versions of disruption gives similar results, $D$ is the only one that is both robust over time, and also predicts delayed citation bursts for articles in the future.

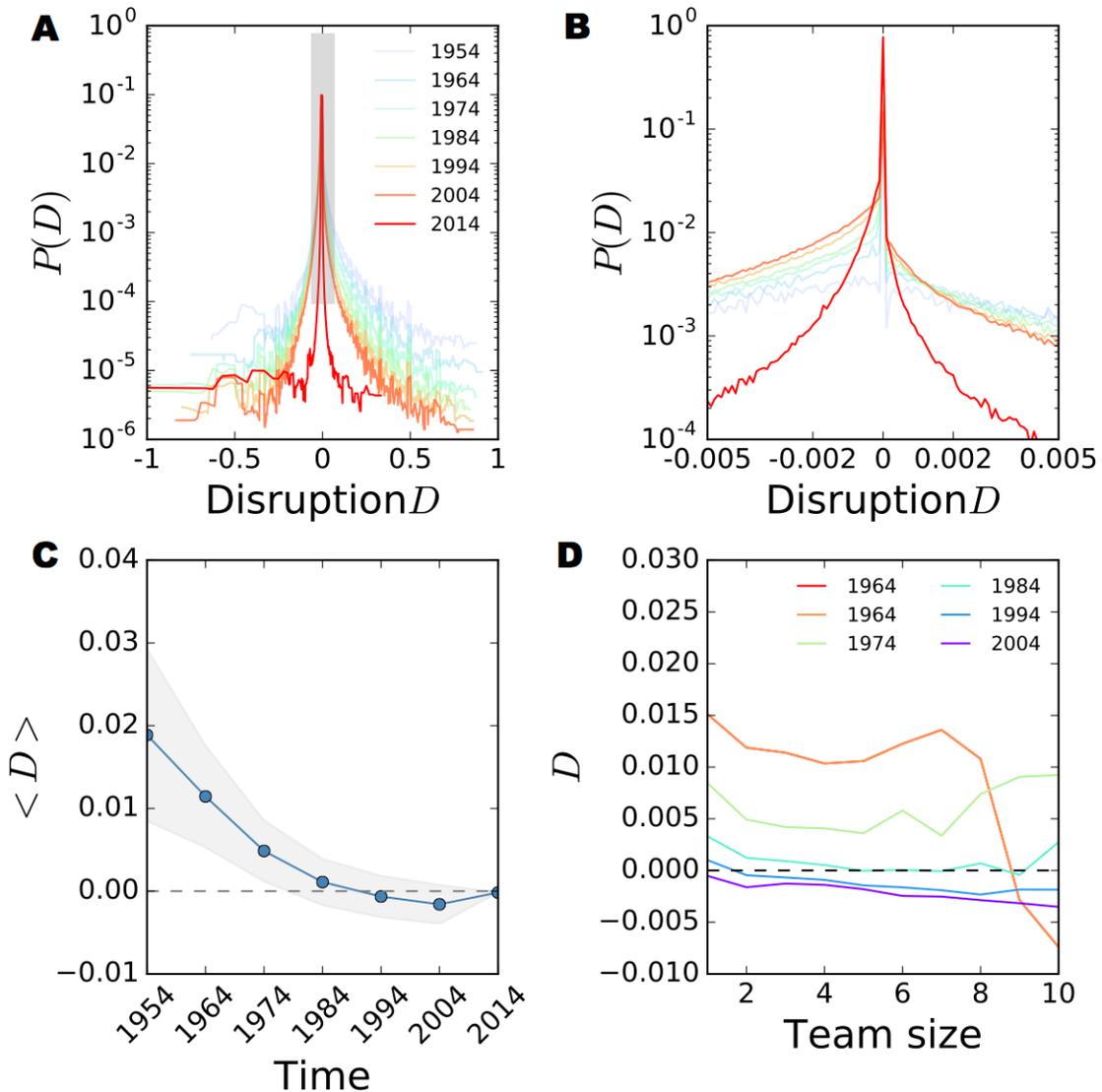

**Figure S4. Temporal evolution of disruption across WOS articles.** (A) The distribution of disruption across six decades. A majority of arcticles have a disruption value close to zero. Earlier cohorts (blue curves) are more disruptive and developmental—they exhibit thicker tails in the disruption distribution—than later cohorts, which face a much deeper collection of potentially relevant prior papers on which to build or ignore. (B) A zoom-in version of the gray region of (A) revealing that later years demonstrate a higher peak at zero. (C) Average disruption across the WOS sample decrease smoothly and consistently over time. (D) The negative correlation between disruption and team size holds across time periods. Unlike the main body of the paper, which renders disruption in terms of percentile change, this is measured in the native metric of disruption to highlight the shift with time. Nevertheless, with changes in team size, each cohort traverses a majority of the total variation of disruption for that cohort.

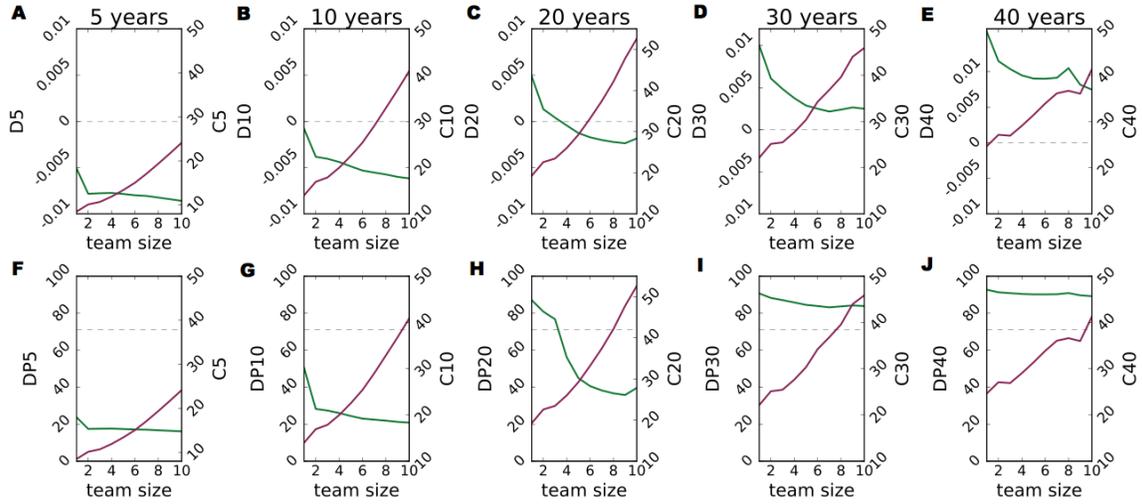

**Figure S5. Decreasing disruption and increasing impact with growing team size are robust to changes in the width of the time-window of observation for articles publised in 1970.** (A) ~ (E) We select the articles published in 1970 and calculate $D(t)$ (the left axis) and $C(t)$ (the right axis) with time-window length $t$ in years, and find that the increase of $D(t)$ and decrease of $C(t)$ with growth of $m$ are robust to widening the time-window, despite the fact that average levels of $C(t)$ and $D(t)$ go with time. In (F) ~ (J) We present similar results as (A) ~ (E), except that we measure total disruption $DP(t)$ instead of $D(t)$. Disruption percentile is calculated based on the distribution of disruption values over all papers for the widest time window (40 years).

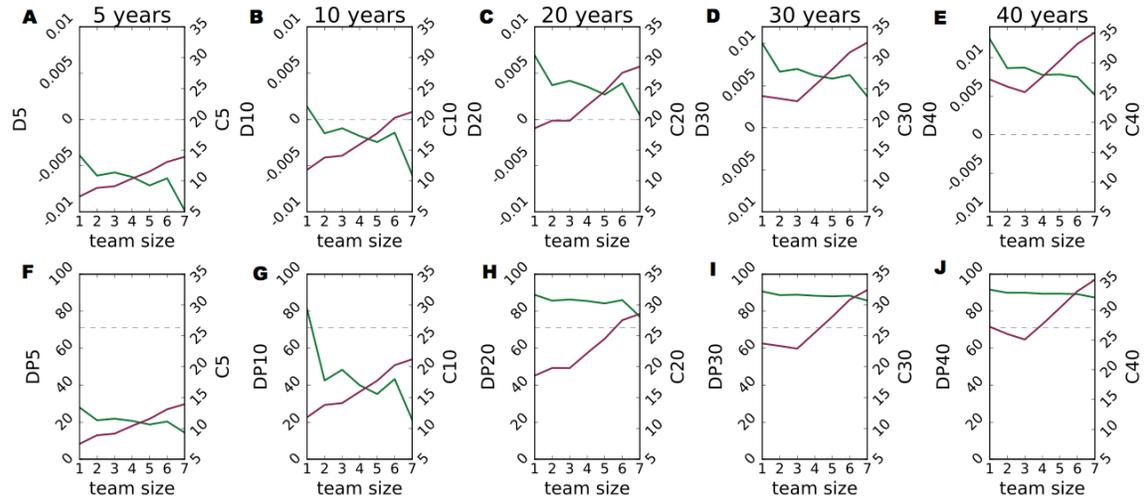

**Figure S6. Decreasing disruption and increasing impact with growing team size are robust to changes in the width of the time-window of observation for all years.** Same as Figure S4 but we use all articles from all years and not those published in a single year. More specifically, we use the papers published from 1954-2009 for impact *C(5)* and disruption *D(5)*, 1954-2004 for *C(10)* and *D(10)*, 1954-1994 for *C(20)* and *D(20)*, 1954-1984 for *C(30)* and *D(30)*, 1954-1974 for *C(40)* and *D(40)*. As observed in Fig. S4, the increase of *D(t)* and decrease of *C(t)* with the increase of *m* are robust to the shifts in time-window width, despite the fact that the average levels of *C(t)* and *D(t)* rise over time.

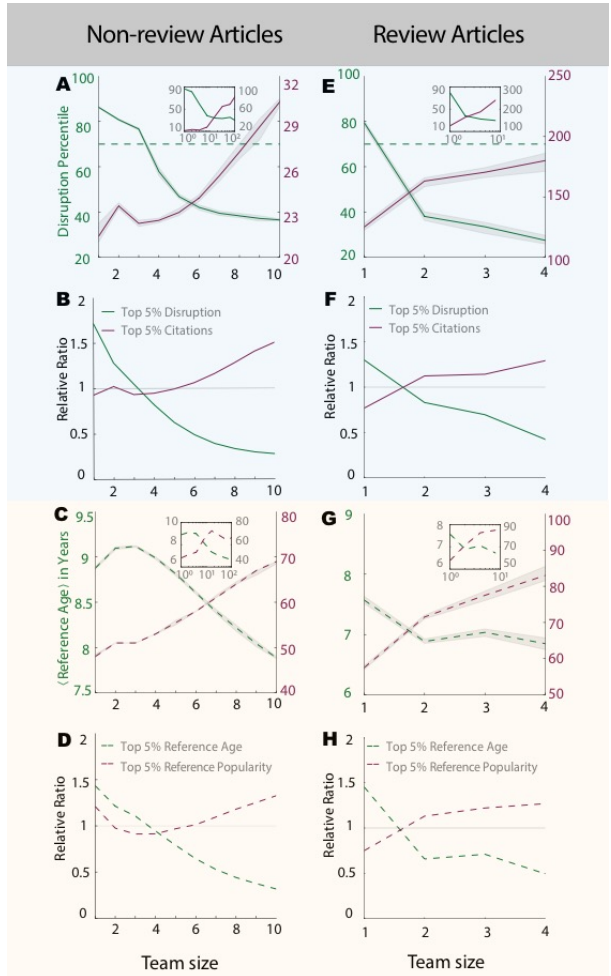

**Figure S7. Comparison between non-review articles (A-D) and review articles (E-H).** We select these articles based on journal title. Among the 15,146 journals in our dataset, 458 of them have the word "review" in the title and the remaining 14,688 do not. To exclude non-review, article journals like Physical Review E, we further select 48 journals that have both "annual" and "review" in the title. By fetching all articles published on the selected journals, we obtain 22,672 review articles and 23,435,943 non-review articles. We investigate the relationship of four variables with team size as in Figure 2, including disruption percentile, number of citations, average age of reference, and median popularity of references. We observe that non-review and review articles show the same dynamics in all variables investigated. Note that as there are almost always less authors in review articles than non-review articles, for reviews we only display team size from one to four as shown in Panels E-H, which covers 97% of the population, and for the non-reviews we display team size from one to ten as shown in Panels A-D, which covers 98% of the population. The range of team size in the two subsets graphed in E and G are naturally smaller than the subsets in A and C.

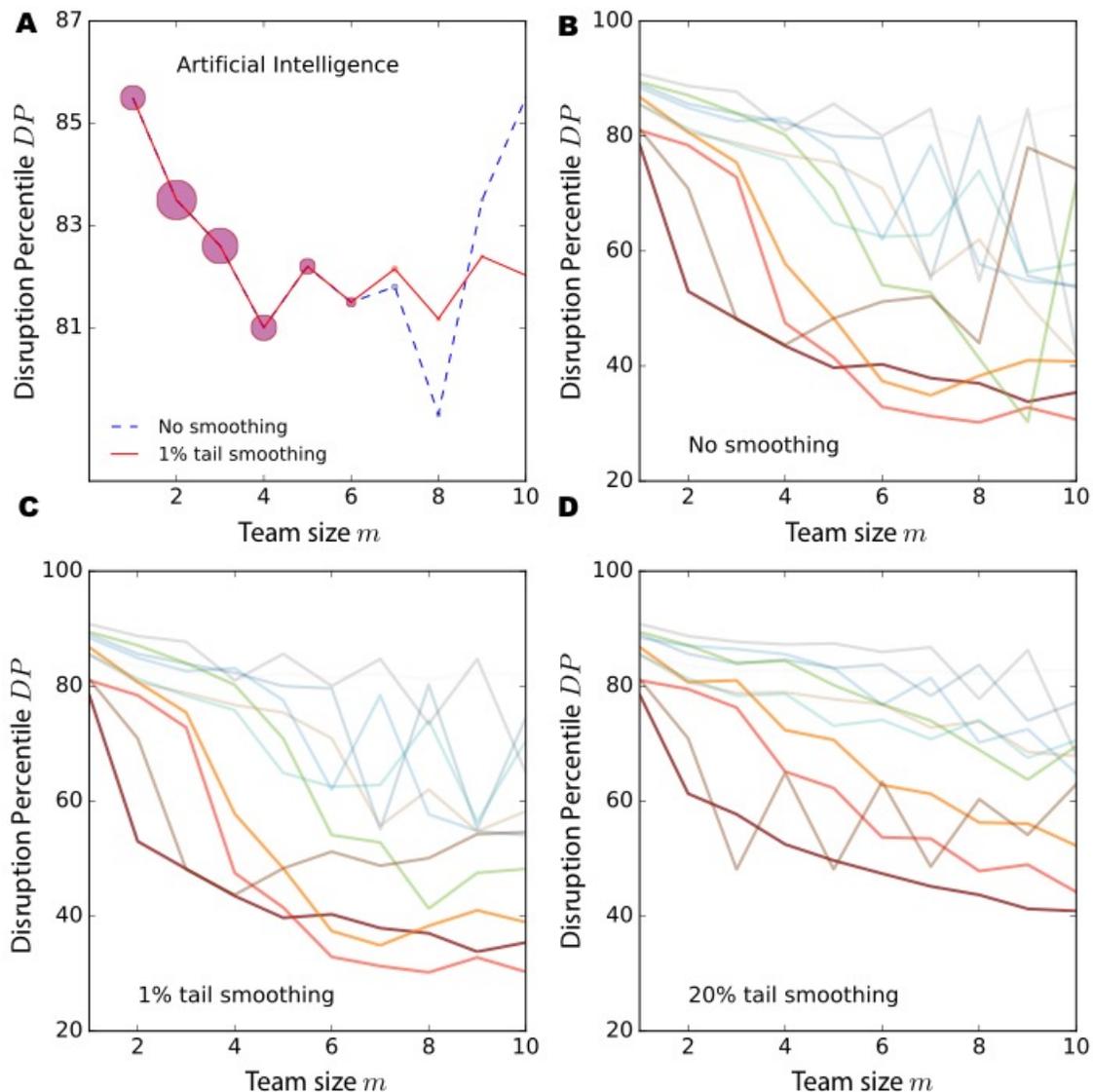

**Figure S8. Weighted moving average technique for data smoothing.** The relationship between team size and disruption may be noisy due to a lack of data when we analyze articles from the same journal or field. For example, less than 1% of articles in "Artificial Intelligence" (a subfield of "Computer and Information Technology") have more than 6 authors, but these articles contribute to substantial variance in the data (A). We use the moving average technique to limit noise in the data. More specifically, we define parameter $k$, which provides threshold value $m_k$ for team size $m$ such that $P(m>m_k) < k$. For any data point with a team size greater than $m_k$, its disruption percentile $DP_m$ will be updated to be the average between its current value and the value of its left neighbor, $DP_{m-1}$, weighted by corresponding sample sizes (the number of articles for a given team size). (A) Curves for the subfield "Artificial Intelligence" before and after smoothing, where circle size is proportional to sample size. (B), (C), and (D) show how the smoothing effect depends on the value of *k* across 13 subfields. We select k=1% to generate Figure 3B in the main text.

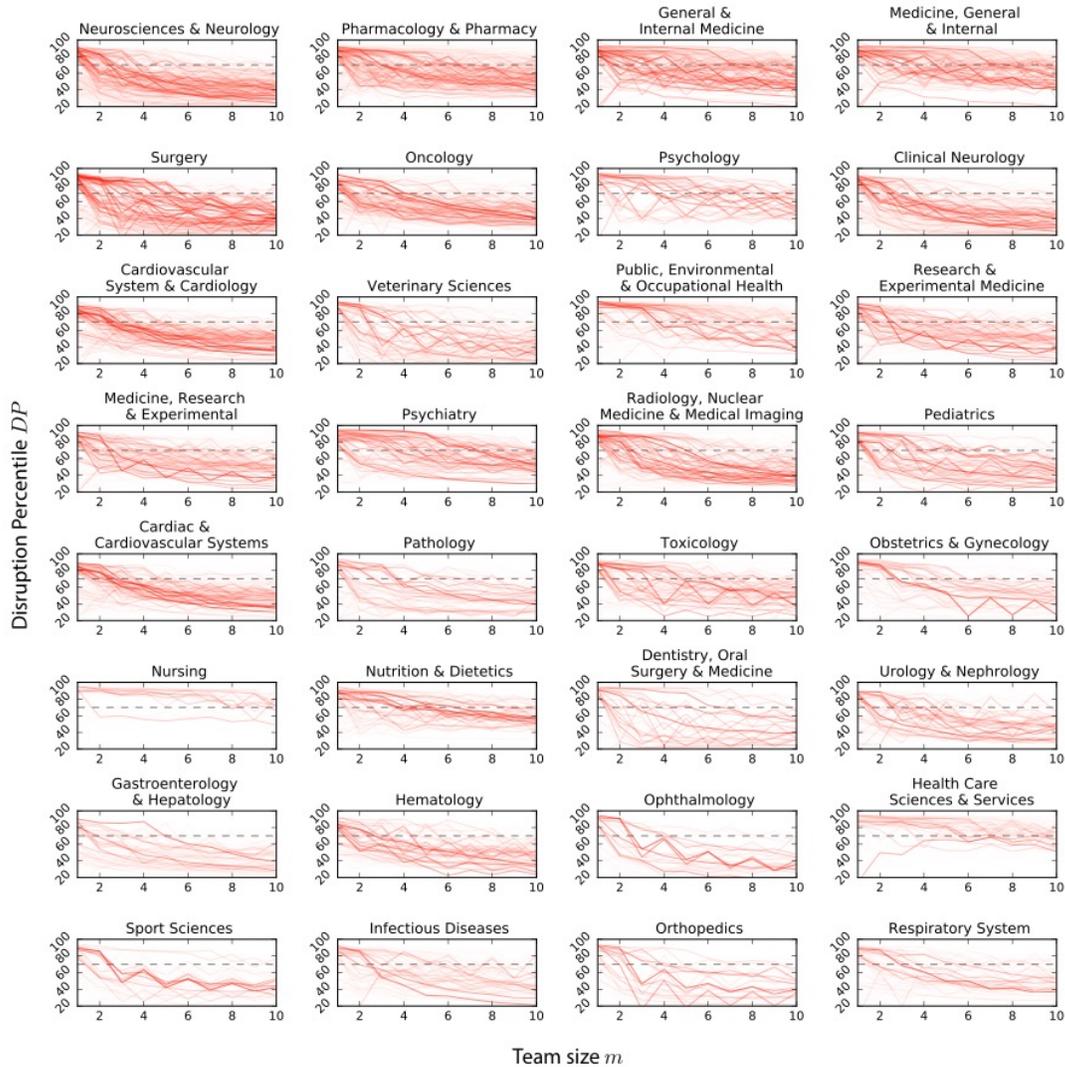

**Figure S9. Disruption percentile vs. team size for Medicine journals.** Each curve corresponds to a journal (only journals with more than three data points are shown) and each panel corresponds to a subfield. Curves are smoothed using the technique introduced in Figure S8, by setting the smoothing parameter k=0.2. Darkness of curves is proportional to both sample size and the absolute value of the regression coefficient of disruption percentile on team size, so that journals with more articles and that display stronger (both negative and positive) relationships are more distinguishable from the background.

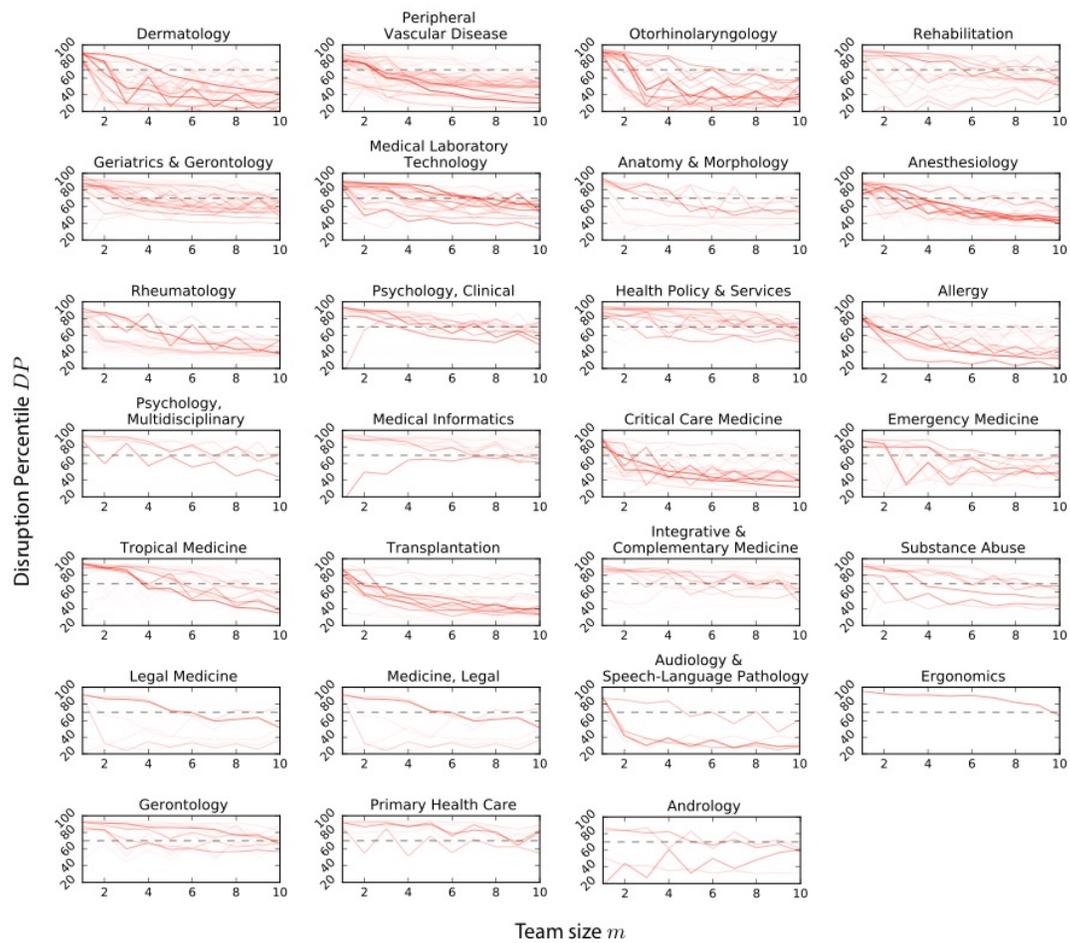

**Figure S10. Disruption percentile vs. team size for Medicine journals.** Each curve corresponds to a journal (only journals with more than three data points are shown) and each panel to a subfield. This figure is created following the same method as Figure S9.

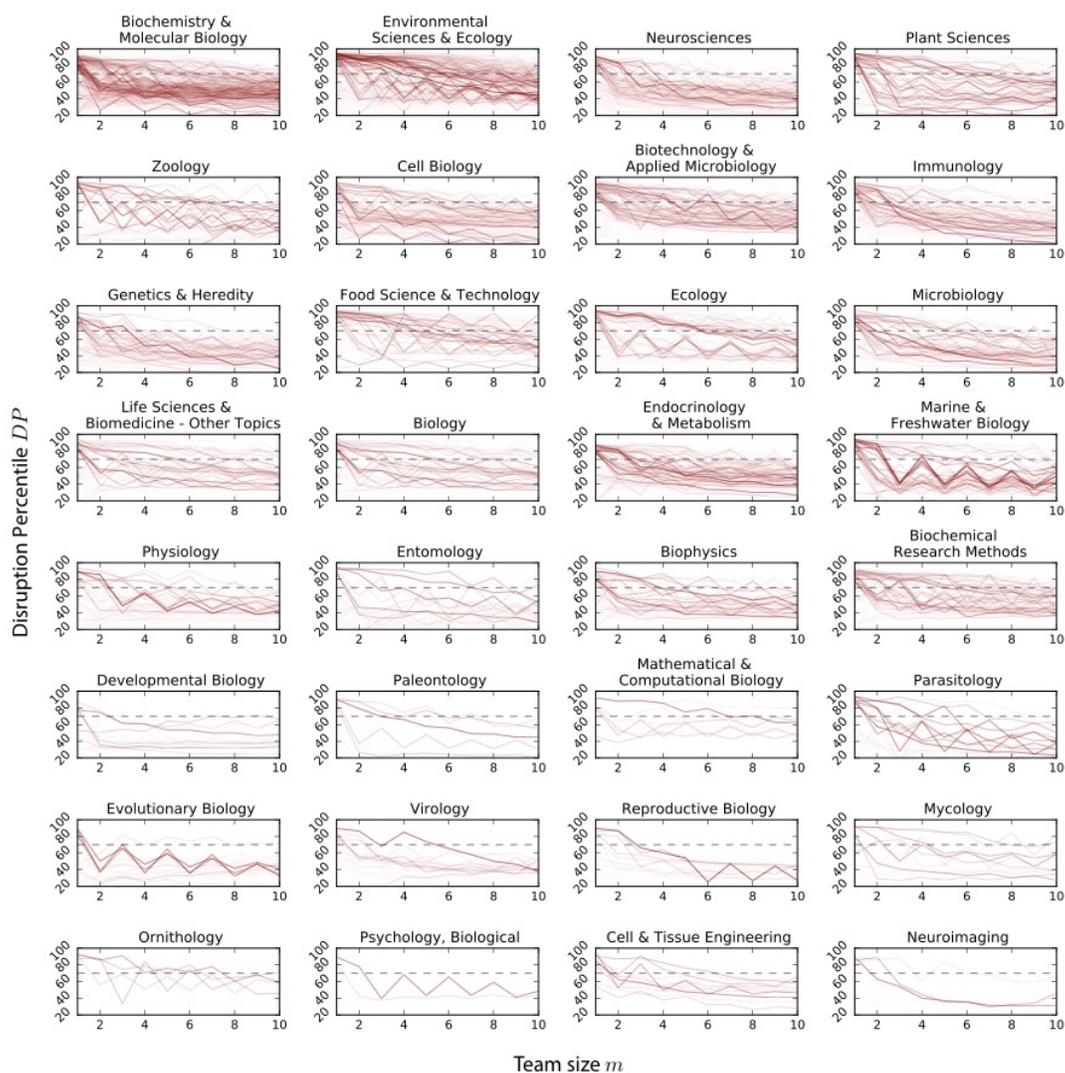

**Figure S11. Disruption percentile vs. team size for Biology journals.** Each curve corresponds to a journal (only journals with more than three data points are shown) and each panel to a subfield. This figure is created following the same method as Figure S9.

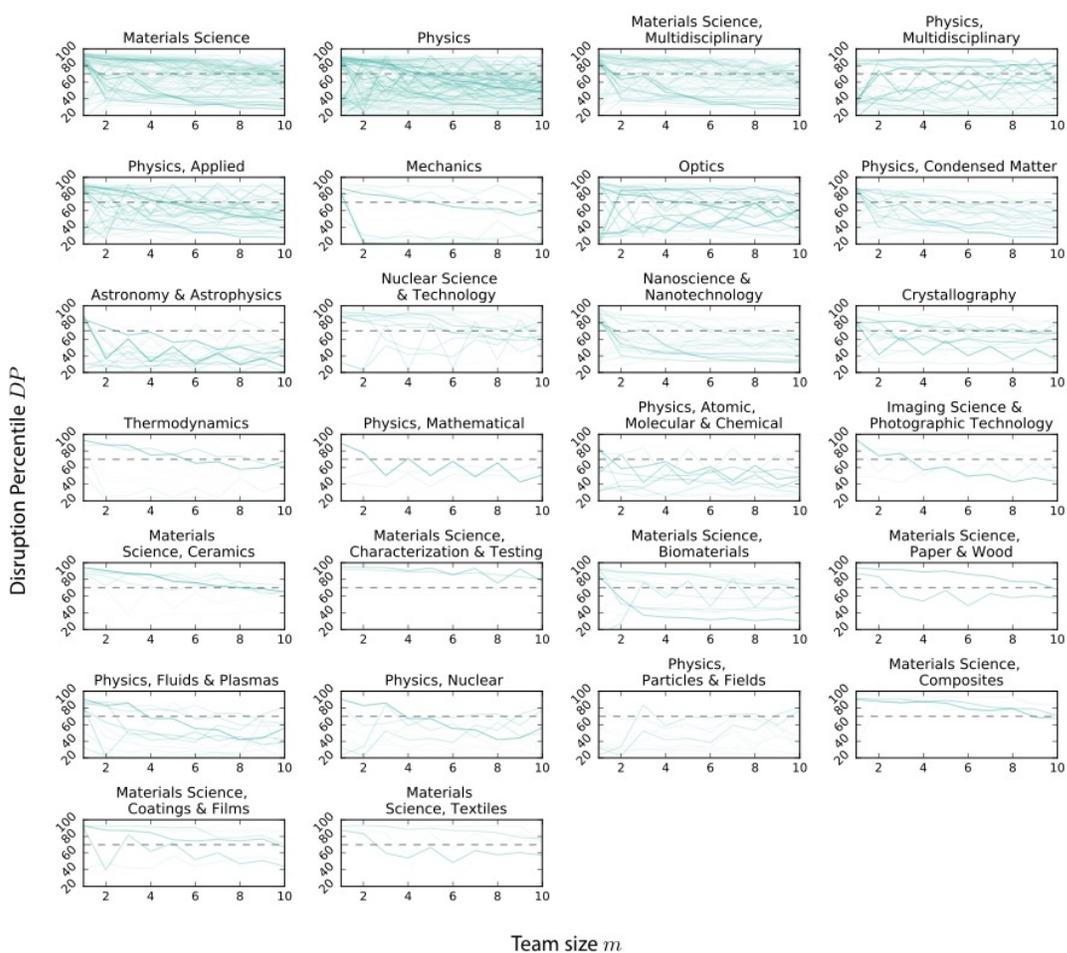

**Figure S12. Disruption percentile vs. team size for Physical Science journals.** Each curve corresponds to a journal (only journals with more than three data points are shown) and each panel to a subfield. This figure is created following the same method as Figure S9.

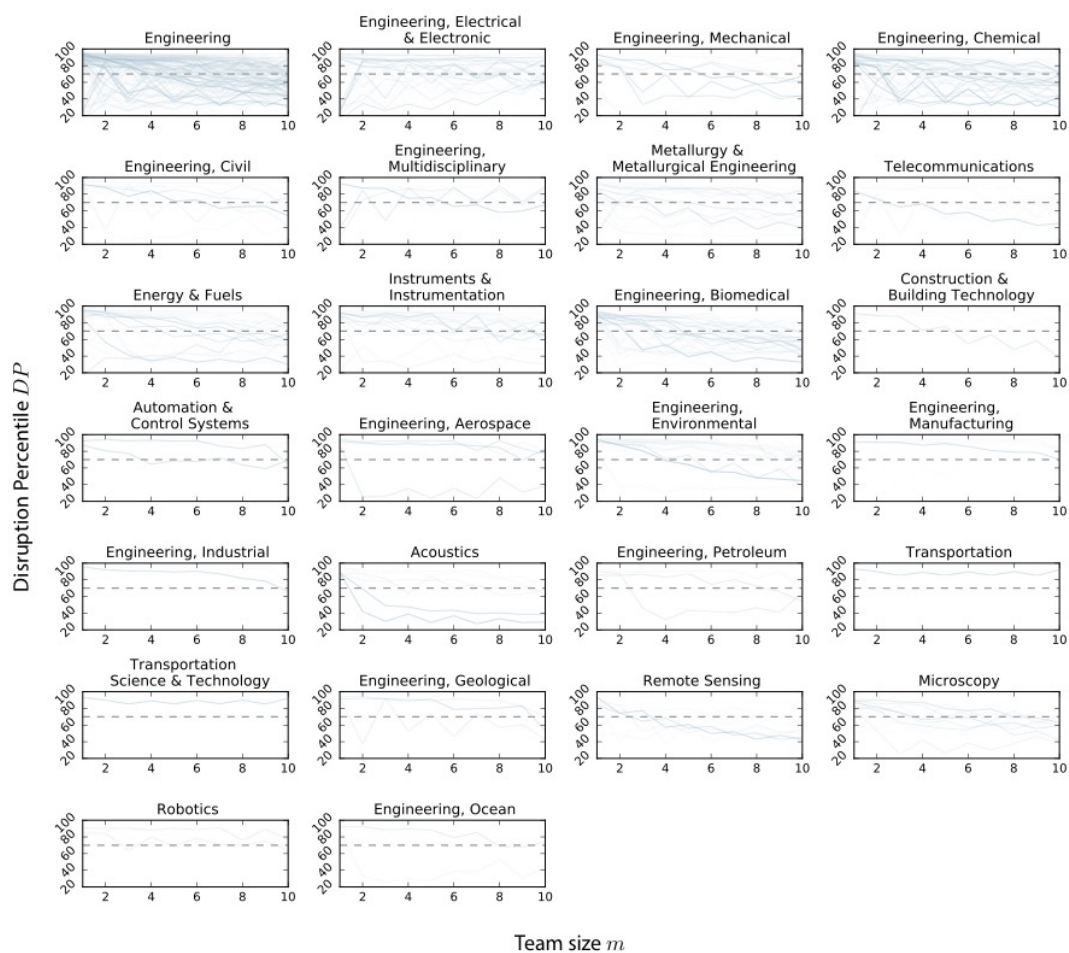

**Figure S13. Disruption percentile vs. team size for Engineering journals.** Each curve corresponds to a journal (only journals with more than three data points are shown) and each panel to a subfield. This figure is created following the same method as Figure S9.

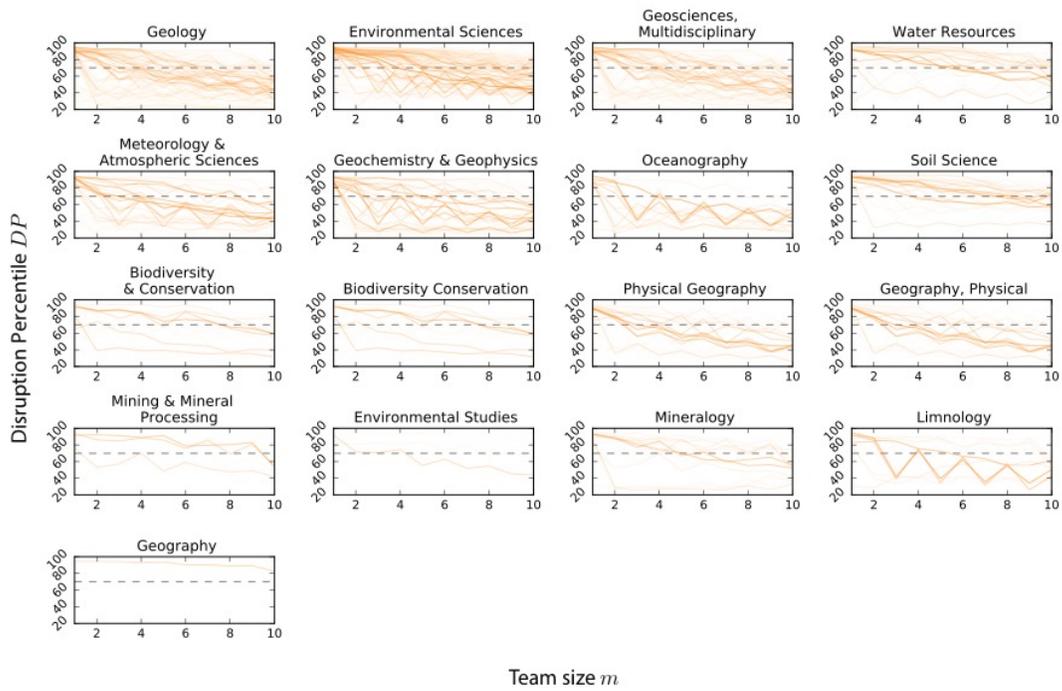

**Figure S14. Disruption percentile vs. team size for Environmental and earth sciences journals.** Each curve corresponds to a journal (only journals with more than three data points are shown) and each panel to a subfield. This figure is created following the same method as Figure S9.

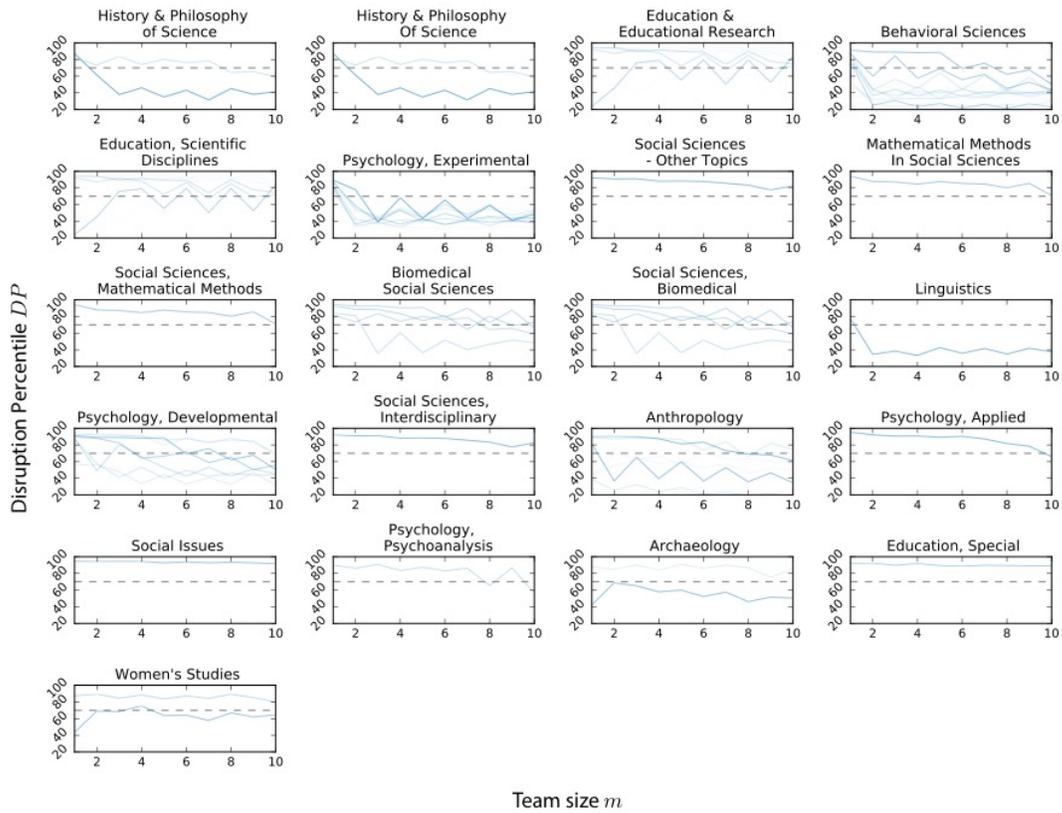

**Figure S15. Disruption percentile vs. team size for Social Sciences journals.** Each curve corresponds to a journal (only journals with more than three data points are shown) and each panel to a subfield. This figure is created following the same method as Figure S9.

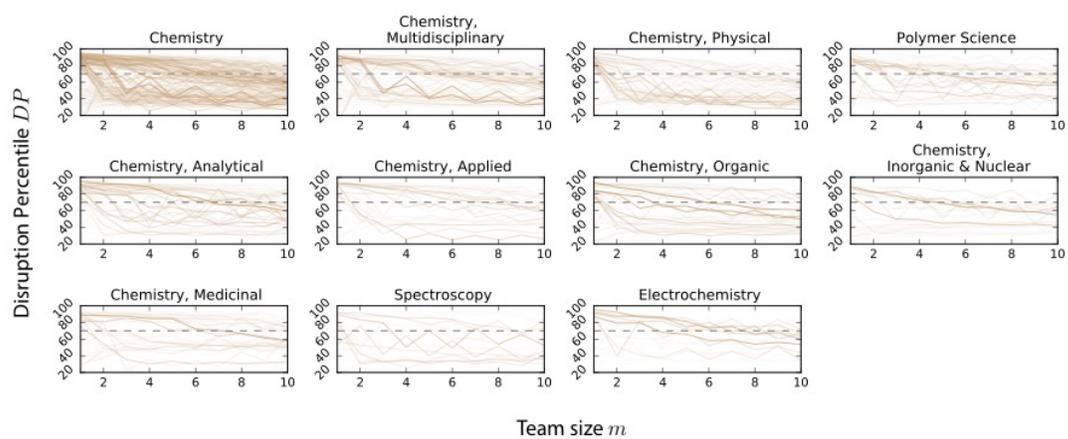

**Figure S16. Disruption percentile vs. team size for Chemistry journals.** Each curve corresponds to a journal (only journals with more than three data points are shown) and each panel to a subfield. This figure is created following the same method as Figure S9.

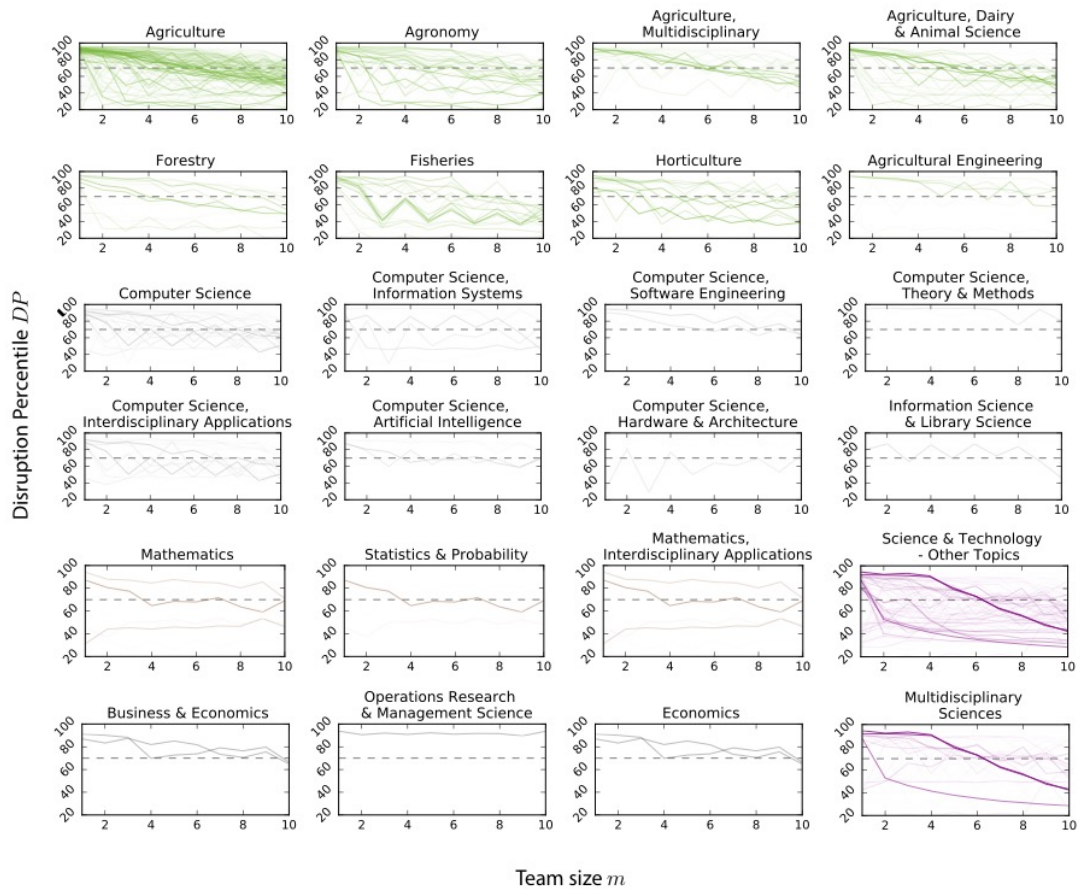

**Figure S17. Disruption percentile vs. team size for Agriculture, Computer and information technology, Business and management, Mathematics, and Multidisciplinary Sciences journals.** Each curve corresponds to a journal (only journals with more than three data points are shown) and each panel to a subfield. This figure is created following the same method as Figure S9.

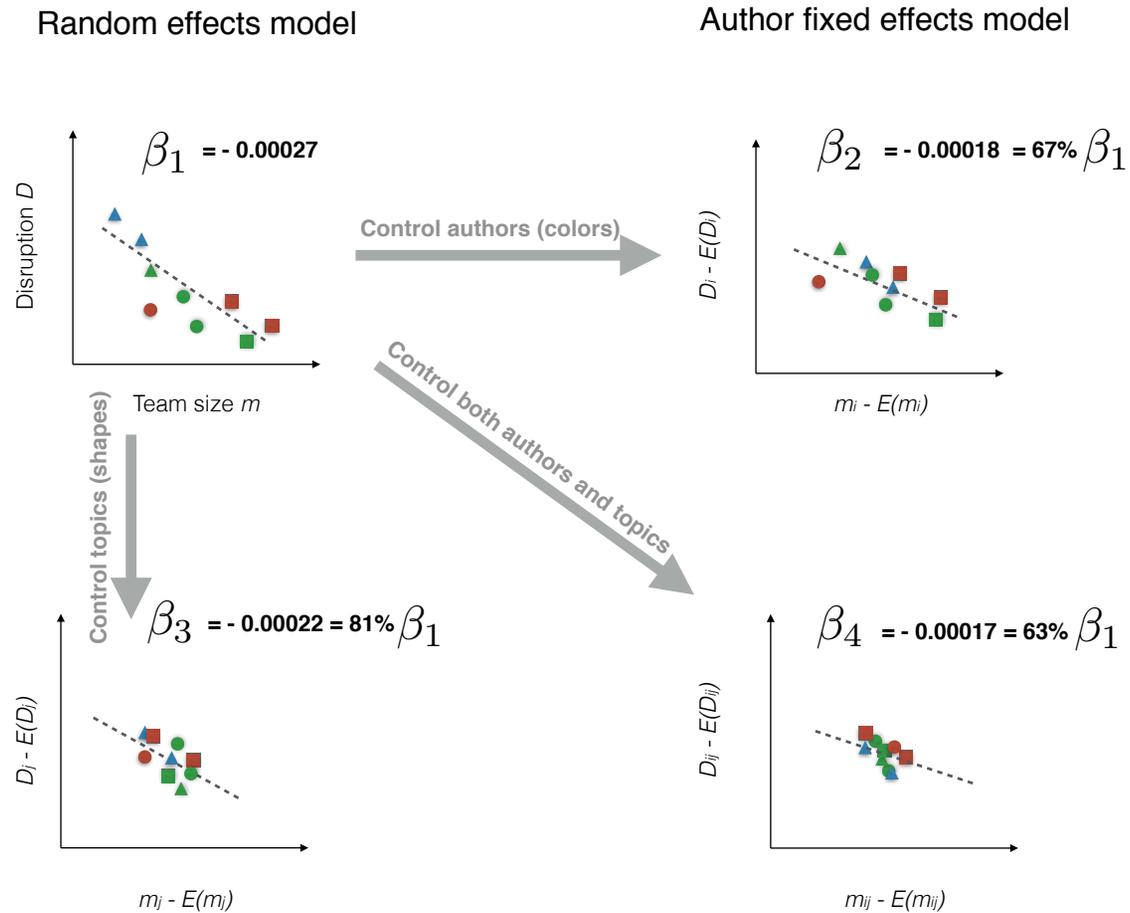

**Figure S18. Diagram illustrating our random and fixed effects strategy for identifying the influence of teams on disruption, controlling for author and topic.** See supplemental text for details. All beta coefficients are significant at $p < 10^{-54}$.

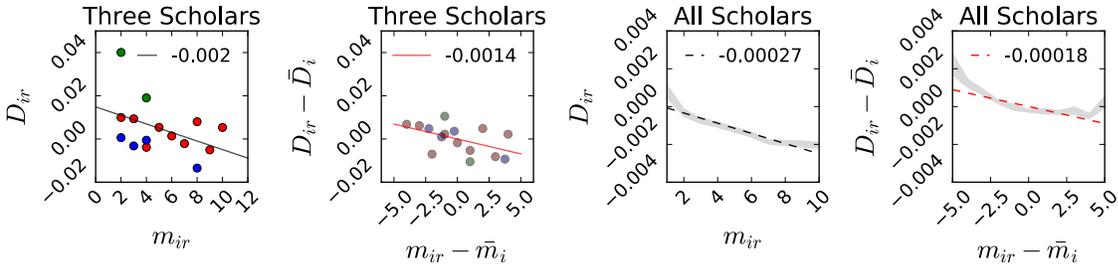

**Figure S19. Examples of random effects and author fixed effects models, which regress disruption on team size.** (A) We construct the random effect model $D_{ir} = \alpha_1 + \beta_1 m_{ir} + \varepsilon_{ir}$ for three selected scholars, in which $D_{ir}$ stands for disruption of the article by the $i$th scholar in the $r$th smallest team he/she participates and $m_{ir}$ represents the size of that team. Article disruption from author's teams of the same size are averaged. $\alpha_1$ and $\beta_1$ are the intercept and team size regression coefficients, respectively. Data from three scholars are graphed as red, blue, and green dots, respectively. The regression coefficient is estimated as $\beta_1 = -0.002$. (B) We construct the fixed effect model $D_{ir} - \overline{D}_i = \alpha_2 + \beta_2 ( m_{ir} - \overline{m}_i ) + \varepsilon_{ir}$ to control for author-specific effects in the relationship of disruption and team size across the three individual scholars and find that the impact of team size on disruption is still significant. The regression coefficient is estimated as $\beta_2 = -0.0014$. (C) The same as (A) but using data from all 88,863 scholars. The gray area graphs bootstrapped 95% confidence-intervals for the population mean. The black dotted line traces the random effects model prediction and the red dotted line follows the author fixed effects model. (D) The slope and confidence interval corresponding to the author fixed effects model.

**Table S1. Example results from the Doc2vec model.**

| | |
|---|---|
| Example Text (Title and Abstract) | constrained optimal discrimination designs for fourier regression models in this article the problem of constructing efficient discrimination designs in fourier regression model is considered we propose designs which maximize the power of the test which discriminates between the two highest order models subject to the constraints that the tests that discriminate between lower order models have at least some given relative power complete solution is presented in terms of the canonical moments of the optimal designs and for the special case of equal constraints even more specific formulae are available |
| Most similar | optimal design for smoothing splines in the common nonparametric regression model we consider the problem of constructing optimal designs if the unknown curve is estimated by smoothing spline special basis for the space of natural splines is introduced and the local minimax property for these splines is used to derive two optimality criteria for the construction of optimal designs the first criterion determines the design for most precise estimation of the coefficients in the spline representation and corresponds to optimality while the second criterion is the optimality criterion and corresponds to an accurate prediction of the curve several properties of the optimal designs are derived in general and optimal designs are not equivalent optimal designs are determined numerically and compared with the uniform design |
| Least similar | layer by layer films of hemoglobin or myoglobin assembled with zeolite particles and positively charged hemoglobin hb or myoglobin mb at ph in solutions and negatively charged zeolite particles in dispersions were alternately adsorbed onto solid surfaces forming zeolite protein layer by layer films which was confirmed by quartz crystal microbalance qcm and cyclic voltammetry cv the protein films assembled on pyrolytic graphite pg electrodes exhibited pair of well defined nearly reversible cv peaks at about vs sce at ph characteristic of the heme fe iii fe ii redox couples hydrogen peroxide and nitrite no in solution were catalytically reduced at zeolite protein film modified electrodes and could be quantitatively determined by cv and amperometry the shape and position of infrared amide and ii bands of hb or mb in zeolite protein films suggest that the proteins retain their near native structure in the films the penetration experiments of fe cn as the electroactive probe into these films and scanning electron microscopy sem results indicate that the films possess great amount of pores or channels the porous structure of zeolite protein films is beneficial to counterion transport which is crucial for protein in films controlled by the charge hopping mechanism and is also helpful for the diffusion of catalysis substrates into the films the proteins with negatively charged net surface charges at ph were also successfully assembled with like charged zeolite particles into layer by layer films although the adsorption amount was less than that assembled at ph the possible reasons for this were discussed and the driving forces were explored elsevier all rights reserved |

**Table S2. Example results of the Doc2vec model.**

| | |
|---|---|
| Example Text (Title and Abstract) | probing the very high redshift universe with gamma ray bursts prospects for observations with future ray instruments gamma ray bursts grbs are the most violent explosions in the universe long duration grbs are associated with the collapse of massive stars rivalling their host galaxies in luminosity the discovery of the most distant confirmed object in the universe grb opened new window on the high redshift universe making it possible to study the cosmic reionization epoch and the preceding dark ages as well as the generation of the first stars population iii using grbs obviously this enables wealth of new studies using the near infrared nir characteristics of grb afterglows here we explore different path focusing on the next generation of ray missions with large area focusing telescopes and fast repointing capabilities we found that ray data can complement nir observations and for the brightest grbs can provide an accurate and independent redshift determination metallicity studies can also be carried out profitably once the redshift is known finally we discuss observational signatures of grbs arising from population iii stars in the ray band |
| Most similar | detection of gamma ray emission from the vela pulsar wind nebula with agile pulsars are known to power winds of relativistic particles that can produce bright nebulae by interacting with the surrounding medium these pulsar wind nebulae are observed by their radio optical and ray emissions and in some cases also at tev teraelectron volt energies but the lack of information in the gamma ray band precludes drawing comprehensive multiwavelength picture of their phenomenology and emission mechanisms using data from the agile satellite we detected the vela pulsar wind nebula in the energy range from mev to gev this result constrains the particle population responsible for the gev emission and establishes class of gamma ray emitters that could account for fraction of the unidentified galactic gamma ray sources |
| Least similar | improvement of thermal bond strength and surface properties of cyclic olefin copolymer coc based microfluidic device using the photo grafting technique this paper reports on the development and application of permanent surface modification technique photo grafting as an improved method for bonding coc topas microfluidic substrates with cover plate without affecting the channel integrity this technique not only helps to increase the bond strength of the original device but also makes the surface hydrophilic which is essential for quick fluid flow while passing analytes through the device the bond strength of the modified and unmodified chips was measured using the tensile and peel tests it was observed that the bond strength of the modified chips has increased approximately times to mpa compared to mpa for the unmodified chip the modified surface was evaluated using ray photoelectron spectroscopy fourier transform infrared spectroscopy and water contact angle measurement the contact angle of the modified surface decreased to degrees from degrees for the untreated substrate scanning electron microscope and confocal microscope examinations of cross sectional profiles of the bonded chips indicated that the integrity of the channel features was successfully preserved elsevier all rights reserved |

# Table S3. Example results of the Doc2vec model.

| | |
|---|---|
| Example Text (Title and Abstract) | interspecific differences in heat exchange rates may affect competition between introduced and native freshwater turtles in the iberian peninsula the red eared slider trachemys scripta elegans is an introduced invasive species that is displacing the endangered native spanish terrapin mauremys leprosa however the nature of competitive interactions is relatively unclear in temperate zones mechanisms for maximizing heat retention could be selectively advantageous for aquatic turtle species since individuals usually lose the heat gained from basking very rapidly when entering the water we hypothesized that interspecific differences in morphology and thus in heating and cooling rates might confer competitive advantages to introduced scripta we compared the surface to volume ratios of both introduced and native turtles basing on biometric measures and their effects on thermal exchange rates scripta showed more rounded shape lower surface to volume ratio and greater thermal inertia what facilitates body heat retention and favors the performance of activities and physiological functions such as foraging or digestion thus aggravating the competition process with native turtles in mediterranean habitats |
| Most similar | feeding status and basking requirements of freshwater turtles in an invasion context behavior and feeding status are strongly related in ectotherms trade off between maintenance of energy balance and digestion efficiency has been recently proposed to affect in these animals on the other hand competition for basking sites has been described between iberian turtles and the introduced red eared slider trachemys scripta elegans scripta negatively interferes with basking behavior of native turtles and benefits from greater capacity to retain body heat which may likely result in advantages for the introduced sliders consequently complex effects and alterations in metabolic rates of native turtles might derive from deficient basking behavior we compared the basking requirements of the endangered native spanish terrapin mauremys leprosa and those of the introduced red eared slider analyzing the upper set point temperature lisp defined as the body temperature at which basking ceased of both native and introduced turtles under feeding and fasting conditions we found higher values of lisp in the native species and reduction of this temperature associated with food deprivation in the two turtle species this adjustment of behavior to the nutritional status found in freshwater turtles suggests that ectotherms benefit from metabolic depression as an adaptive mechanism to preserve energy during periods of fasting however reduction in metabolic rates induced by competition with sliders might lead leprosa to prolonged deficiency of their physiological functions thus incurring increased predation risk and health costs and ultimately favoring the recession of this native species in mediterranean habitats elsevier inc all rights reserved |
| Least similar | genetic approaches to assessing evidence for helper type cytokine defect in adult asthma recent evidence suggests that deficiency in the th cytokine pathway may underlie the susceptibility to allergic asthma this study examined whether single nucleotide polymorphisms exist in the promoter region of the two interleukin il subunit genes in patients with asthma messenger rna and protein expressions of signal transducers and activators of transcription il ifn gamma and their receptors are altered in asthma and linkage to genes in the th pathway is present in families with asthma in iceland the promoter regions of the il subunit genes were sequenced in patients with asthma and control subjects without asthma linkage was examined in families that included over patients with asthma and of their unaffected relatives the results demonstrate no evidence of linkage to microsatellite markers in close association with genes within the th pathway and no polymorphism was detected in the promoter regions of the two il subunit genes in the cohort with asthma patients moreover we found no differences in the messenger rna or protein expression signals of genes in the il pathway between the patients and control subjects we conclude that decrease in th type cytokine response is unlikely to present primary event in asthma |

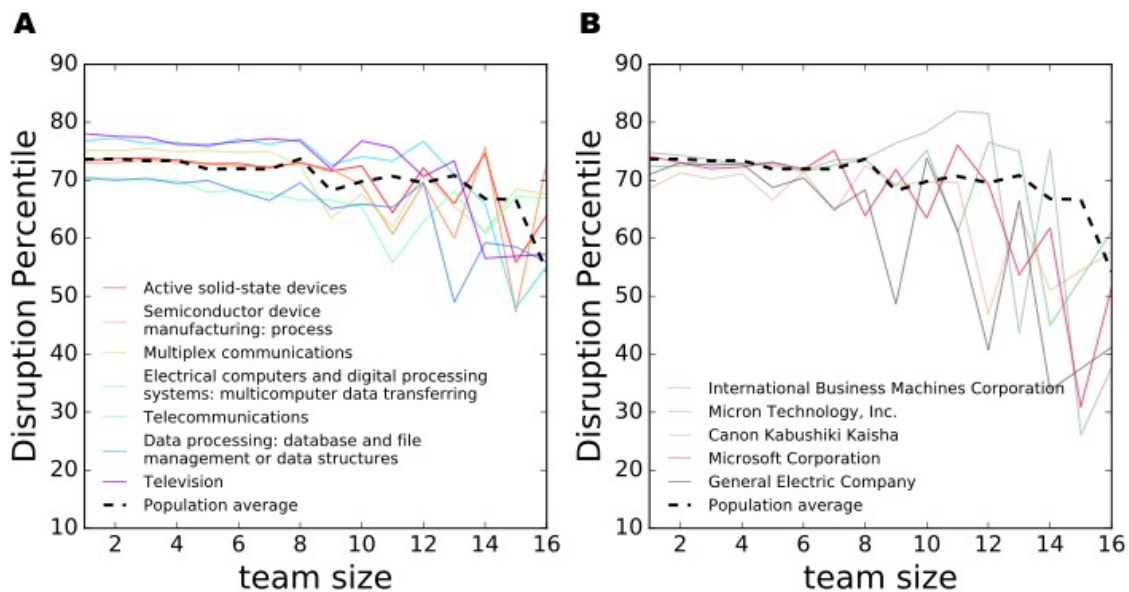

**Figure S20. Disruption percentile vs. team size for U.S. patents across classes and recipients (assignees).** We plot disruption percentile against team size for the top seven classes (A) and the top five recipients (B) against the population average using data from 2002 to 2009. It is observed that the decrease of disruption and increase in team size holds broadly across classes and assignees. The moving average technique introduced in Figure S8 is used to smooth the curve (smoothing parameter k=0.1). As sample size decreases fast with team size in the patent data, we assigned equal weights across team sizes in applying the smoothing technique.

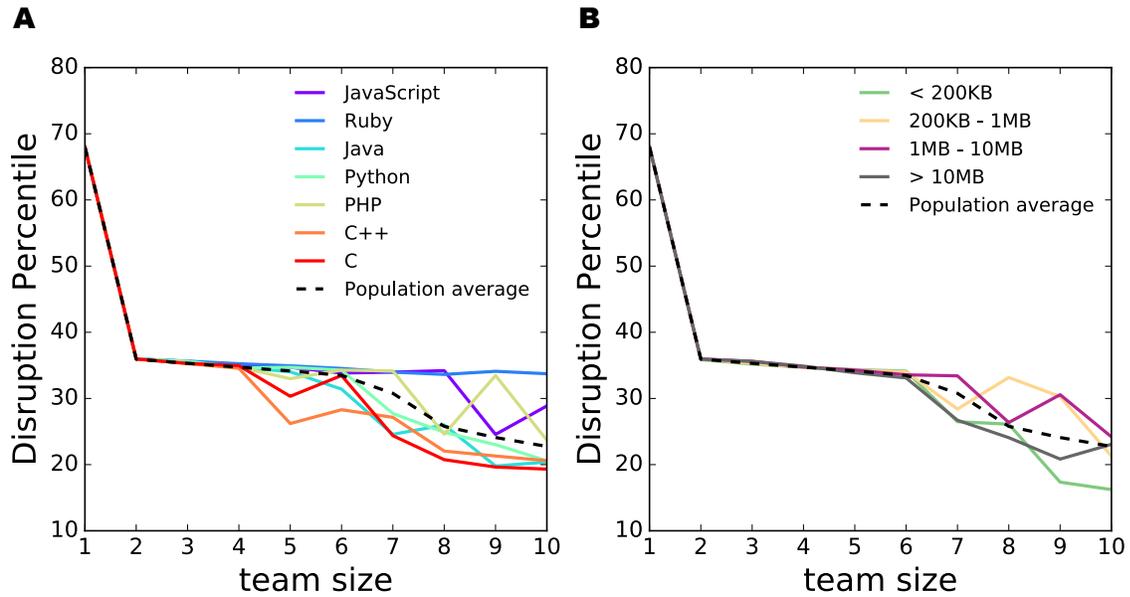

**Figure S21. Disruption percentile vs. team size for GitHub software projects across programming languages and code base sizes.** We plot disruption percentile against team size for the seven most popular programming languages (A) and four scales of code base sizes (B) against the population average using data from 2011-2014. It is observed that the decrease of disruption and increase in team size holds broadly across programming languages and code base sizes.

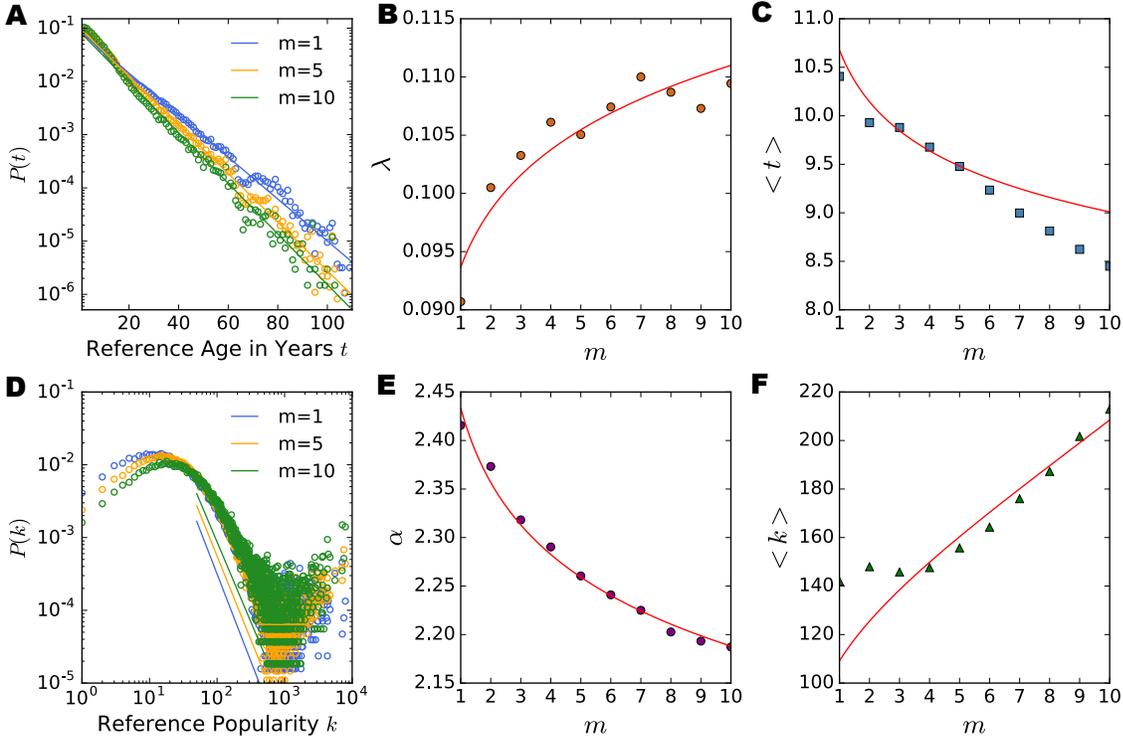

**Figure S22. Reference age decreases with team size, while reference popularity increases with team size in WOS articles.** (A) For articles published in 2010, the probability of observing reference $j$ with age $t$ decreases exponentially with $t$, such that $P(t) \sim e^{\lambda t}$. For larger teams $P(t)$ decreases faster with $t$, suggesting that $\lambda$ is determined by team size $m$. (B) The relationship between $m$ and $\lambda$ (orange circles) can be modeled as $\lambda \sim m^{0.07}$ (red curve). (C) From (A) and (B) we derive the expected value of $t$ by integrating $P(t)$ from zero to maximum $t$, which gives $1/\lambda$, or $m^{-0.07}$. Therefore the expected value of reference age decreases with $m$, as given by blue squares (data) and the red curve (model). (D) The probability of observing reference $j$ with impact $k$ (measured by number of citations) decreases with $k$, following a power-law exponent $\alpha$, supporting the relationship $p(k) \sim k^{-\alpha}$. To eliminate the influence of time we only investigate 5-year old references (published in 2005). We find that larger teams $p(k)$ decrease more slowly with $k$, suggesting that $\alpha$ is determined by $m$. (E) The relationship between $m$ and $\alpha$ (purple circles) can be modeled as $\alpha \sim 2.4m^{-0.05}$ (red curve). (F) From (D) and (E) we derive the expected value of $k$ by integrating $P(k)$ from one to max $k$, which gives $(\alpha-1)/(\alpha-2)$, or $(2.4m^{-0.05} - 1)/(2.4m^{-0.05} - 2)$, as represented by green triangles (data) and the red curve (model).

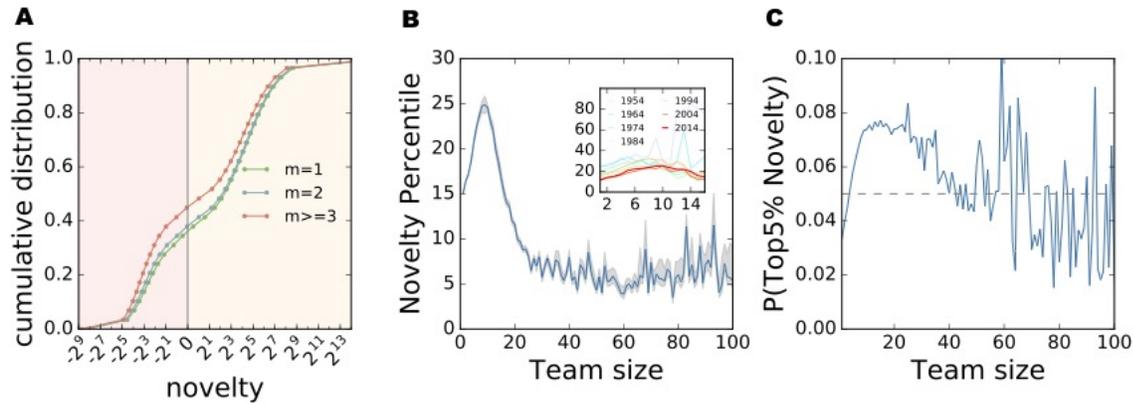

**Figure S23. Changes in combinatorial novelty with team size from WOS articles.** (A) We calculate the pairwise combinational novelty of journals in an article's references using WOS data to replicate their Fig. 3A (33). (B) Uzzi's novelty measure is computed as the tenth percentile value of z-scores for the likelihood that reference sources combine and so the lower value of Uzzi's index indicates higher novelty (33). Here we take the reciprocal of this index and convert it into percentiles to improve readability such that a higher score indicates greater novelty. The gray zone shows the bootstrapped 95% confidence-interval for the population mean. The inset shows the evolution of novelty vs. team size over time. It seems natural that a larger team would provide access to a wider span of literature. Nevertheless, we find that novelty increases with team size $m$, but with diminishing marginal increases to novelty with each additional team member, from $m$=1 to $m$=8. Beyond team of size 8, novelty decreases sharply, reaching its lowest point, at one third of the novelty performed by solo teams at $m$=20 and remaining low and stable thereafter. (C) The probability of observing papers within Top 5% novelty increases with team sizes at first and then decreases. The dotted line shows the the null model that the probability of high novelty is invariant to team size.

9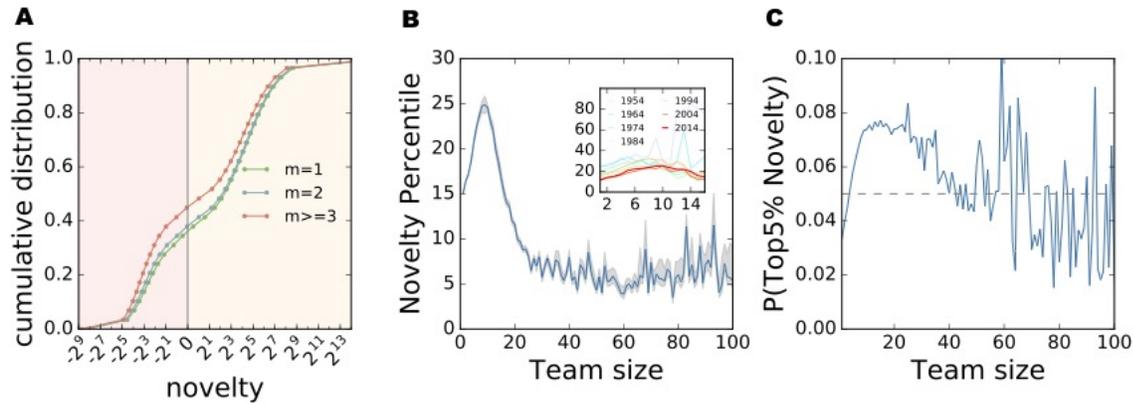

**Figure S23. Changes in combinatorial novelty with team size from WOS articles.** (A) We calculate the pairwise combinational novelty of journals in an article's references using WOS data to replicate their Fig. 3A (33). (B) Uzzi's novelty measure is computed as the tenth percentile value of z-scores for the likelihood that reference sources combine and so the lower value of Uzzi's index indicates higher novelty (33). Here we take the reciprocal of this index and convert it into percentiles to improve readability such that a higher score indicates greater novelty. The gray zone shows the bootstrapped 95% confidence-interval for the population mean. The inset shows the evolution of novelty vs. team size over time. It seems natural that a larger team would provide access to a wider span of literature. Nevertheless, we find that novelty increases with team size $m$, but with diminishing marginal increases to novelty with each additional team member, from $m$=1 to $m$=8. Beyond team of size 8, novelty decreases sharply, reaching its lowest point, at one third of the novelty performed by solo teams at $m$=20 and remaining low and stable thereafter. (C) The probability of observing papers within Top 5% novelty increases with team sizes at first and then decreases. The dotted line shows the the null model that the probability of high novelty is invariant to team size.

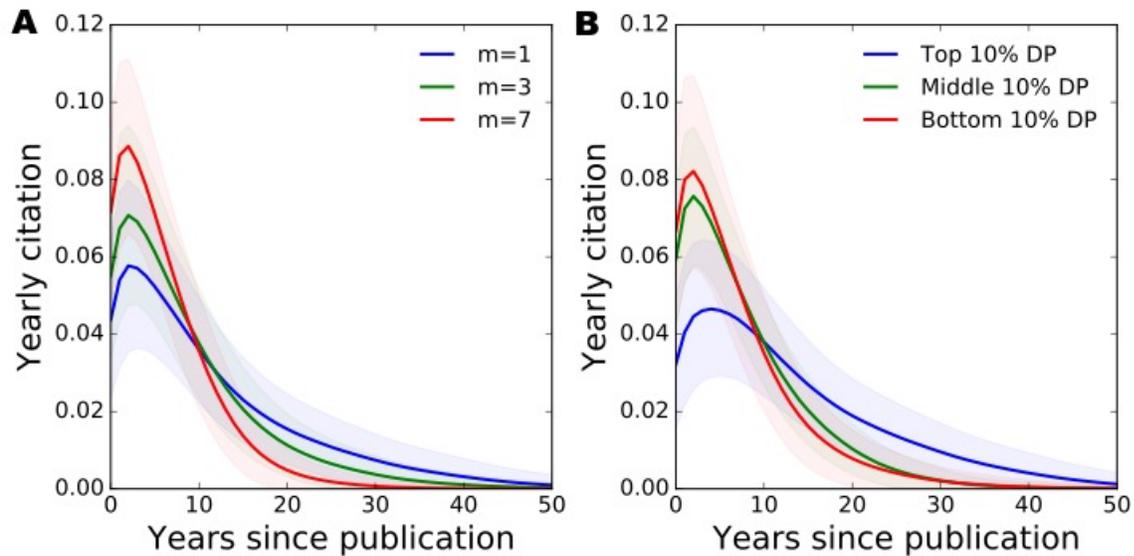

**Figure S24. Comparison of citation dynamics across team sizes and levels of disruption.** To control the difference in temporal citation dynamics across impact levels, we selected 95,474 papers with between 200 and 300 citations from 1954-2014. For each paper we obtain a time series of the number of yearly citations. We then group these papers to calculate their average yearly citations over time. (A) Articles from small teams have a longer season of impact those from big teams. Here *m*=1 (blue curve, 18,532 papers), *m*=3 (green curve, 25,403 papers), and *m*=7 (red curve, 7128 papers) correspond to 10, 50, and 90 percentiles of team size, respectively. (B) Disruptive articles have a longer impact than developmental papers. Here red (37,805 papers), green (4,931 papers), and blue (26,698 papers) curves indicate 0-10, 55-65, and 90-100 percentiles of disruption, respectively. In both panels, curves are smoothed by running average with a time window of five years. Gray regions show one standard deviation of those averages.

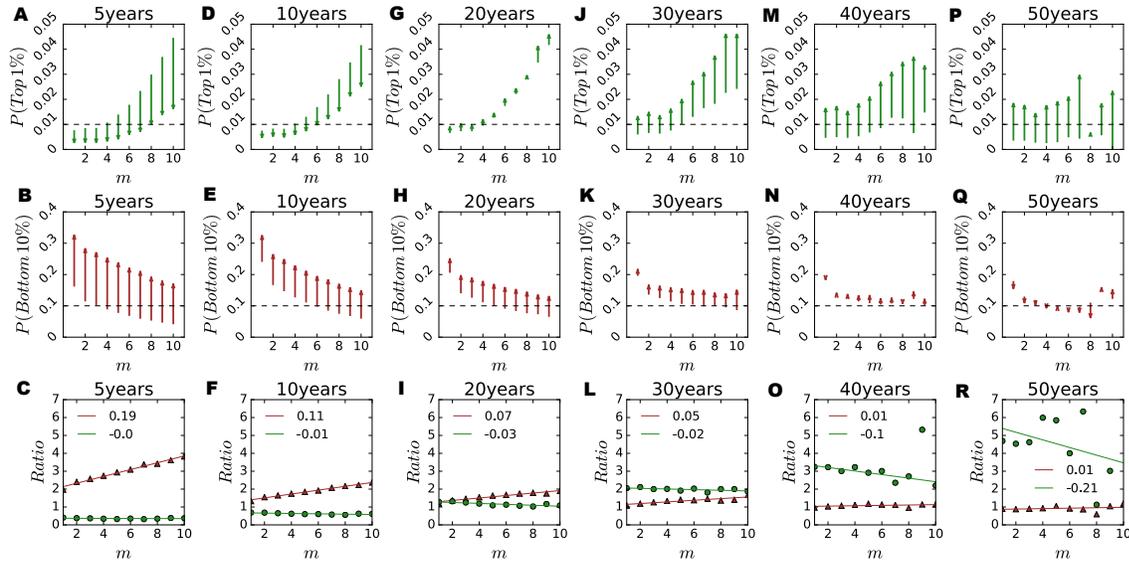

**Figure S25. Temporal risk-reward trade-off in number of citations for WOS articles as a function of different levels of disruption.** (A) The probability of observing articles among the Top 1% most cited ($C(5) > 214$), P(*Top 1%*), is always higher in developmental papers ($D(5) \leq 0$, the origin of green arrows) than disruptive papers ($D(5) > 0$, the target of green arrows) across team sizes. The dotted line indexes the null model that the studied probability is invariant to team size. (B) The probability of observing articles among the Bottom or Tail 10% least cited ($C(5) = 1$), P(*Bottom 10%*), is alway lower in developmental papers ($D(5) \leq 0$, the origin of red arrows) than disruptive papers ($D(5) > 0$, the target of red arrows) across team sizes. The dotted line traces the null model that the studied probability is invariant to team size. (C) Marginal change of probability across team sizes, calculated by dividing the studied Top 1% (green circles) or Bottom 10% (red triangles) probability in developmental papers by its counterpart in disruptive papers. We find that the magnitude of change in P(*Top 1%*) is greater for small teams (regression coefficient < 0) and that in P(*Bottom 10%*) is greater for large teams (regression coefficient > 0). (D) ~ (R) replicate the analysis in (A) ~ (C) but extend the length of the time-window: (D) ~ (F) for 10 years, (G) ~ (I) for 20 years, (J) ~ (L) for 30 years, (M) ~ (O) for 40 years, and (P) ~ (R) for 50 years. In sum, 1) being disruptive increases both risk (P(*Bottom 10%*)) and reward (P(*Top 1%*)) in terms of citation impact; 2) the risk appears immediately and gradually disappears over the long term (≥40 years), whereas the return appears only in the long term (≥20 years); 3) the overtime change in return is more pronounced for small teams and the overtime change in risk is stronger for large teams.

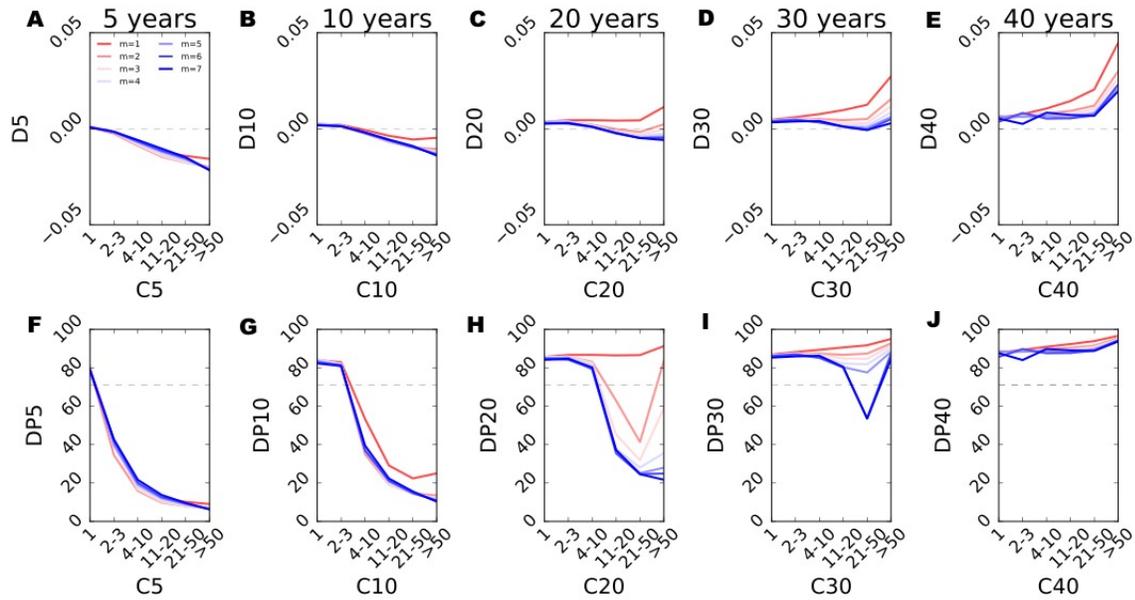

**Figure S26. The negative correlation between disruption and impact in the short term ($t \leq 10$ years) turns positive in the long term ($t >10$ years) for articles published in 1970.** (A) ~ (E) We select articles published in 1970 and calculate $D(t)$ and $C(t)$ with time-window width $t$ in years, and find that they post a negative relationship for 5-year and 10-year windows, but a positive one for 20, 30, and 40 year windows. In (F) ~ (J) we present similar results to (A) ~ (E), except that we portray disruption percentile $DP(t)$ instead of $D(t)$. The percentile is calculated based on the distribution of disruption values of all papers over the longest time window (40 years).

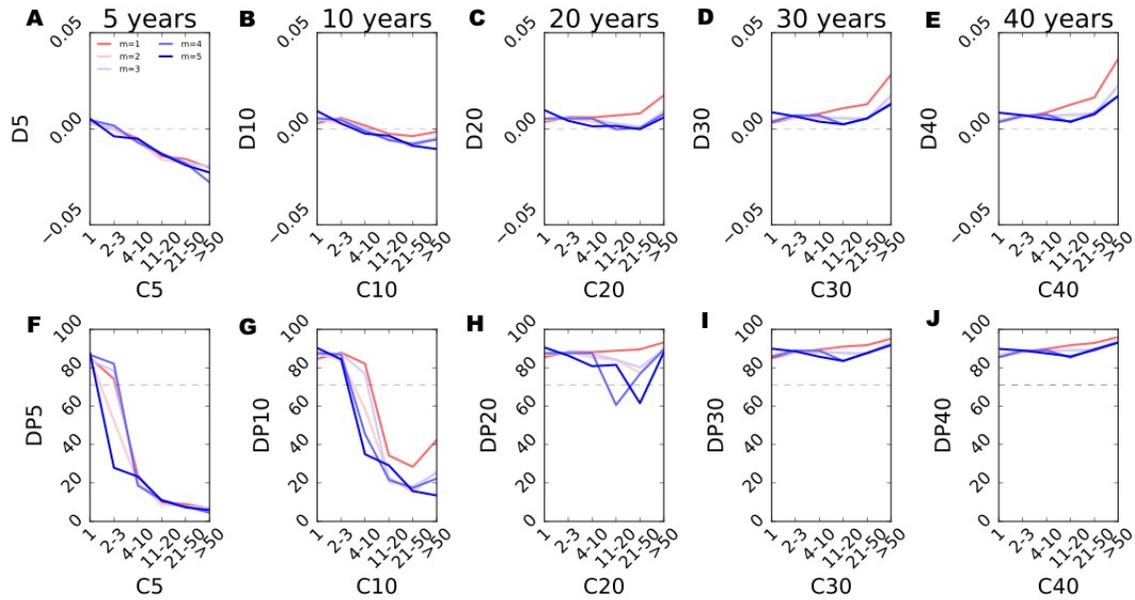

**Figure S27. The negative correlation between disruption and impact in the short term ($t \leq 10$ years) turns positive in the long term ($t >10$ years) for articles published across all years.** Same as Fig. S6 but we use all articles and not a single year. Specifically, we use papers published from 1954-2009 for $C(5)$ and $D(5)$, 1954-2004 for $C(10)$ and $D(10)$, 1954-1994 for $C(20)$ and $D(20)$, 1954-1984 for $C(30)$ and $D(30)$, 1954-1974 for $C(40)$ and $D(40)$. Again, we find that $D(t)$ and $C(t)$ exhibit negative relationship for 5-year and 10-year windows, and they show positive relationship for 20, 30, and 40 years windows, confirming the observation shown in from Figure S6.

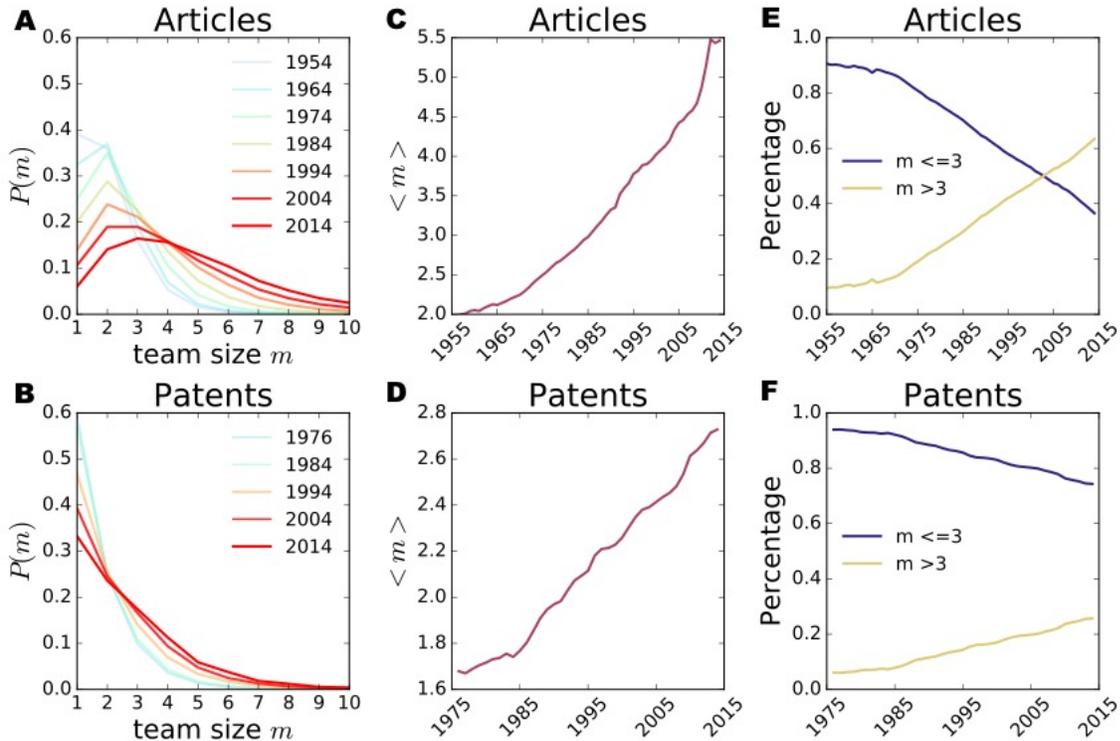

**Figure S28. The decline of small teams.** (A) ~ (B) The evolution of team size distributions over time, for articles and patents, respectively. As time passes, the distribution skews towards larger teams. (C) ~ (D) The increase of average team size $\overline{m}$ over time. The average team size of articles increased from 2 to 5.5 between 1954 and 2014, and that for patents increased from 1.7 to 2.7 between 1976 and 2014. (E) ~ (F) The percentage of small teams ($m \leq 3$) decreased from 91% to 37% in papers and from 94% to 74% for patents during the periods of observation.

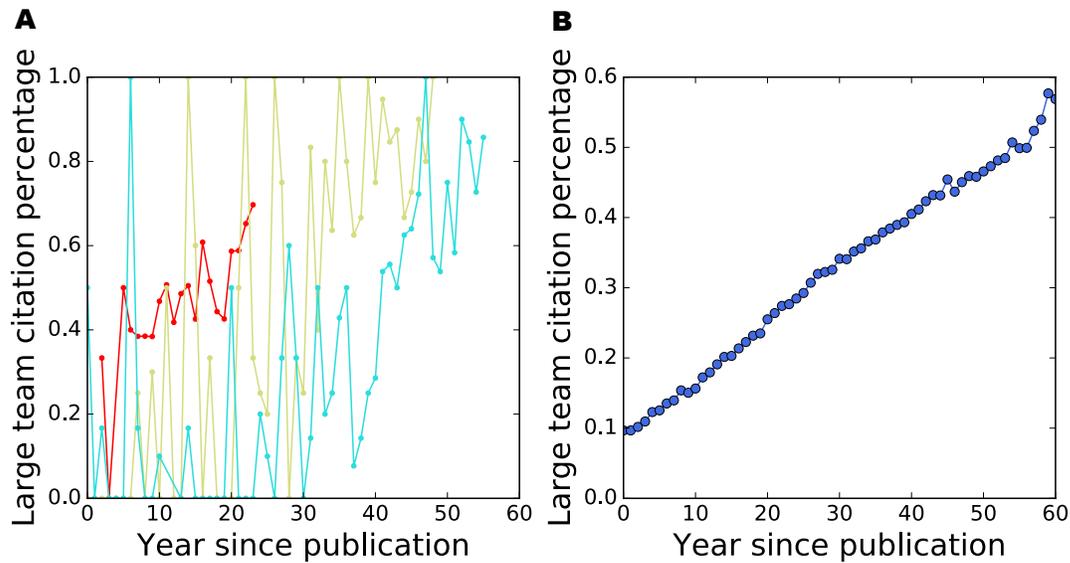

**Figure S29. The ripple effect.** We select 2640 articles that published after 1953 from the WOS dataset that satisfy the following conditions 1) the number of authors is equal to or less than three; 2) the "sleeping beauty" index is among the top 1% of all papers; and 3) the total number of citations in the dataset is among the top 1% of all papers. In other words, these are "small-team sleeping beauties" that have a big impact on science. We trace the citations to these articles and find that the fraction of large teams (with more than three authors) that cite them monotonically and dramatically increases over time. (A) Fraction of citations that come from large teams over time for three example papers. (B) Fraction of large team citation over time, averaged over all the 2640 articles, which attract citations from 657,946 articles contribued by large teams in total.

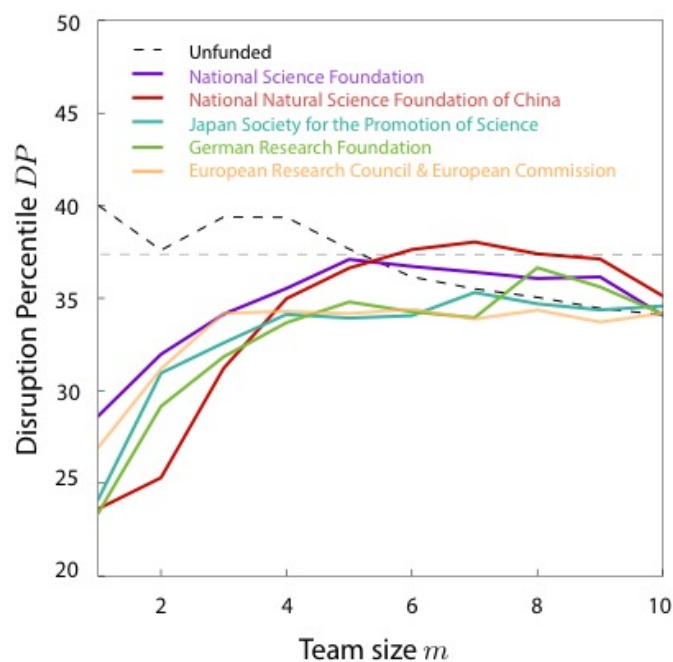

**Figure S30. Underfunded small-team, disruptive research.** Disruption percentile *DP* vs. team size for papers either unfounded or founded by the largest government agencies around the world, including NSF (National Science Foundation), NSFC (National Naturla Science Foundation of China), ERC&EC (Eurpean Research Council and European Commission), DFG (German Research Foundation), and JSPS (Japan Society for the Promotion of Science). The analyzed 4,882,868 papers cover the five time period 2008 - 2012, including 2,216,613 unfunded papers and 2,666,255 founded papers (inlcuing 199,257 for NSF, 55,978 for NSFC, 55,247 ERC&EC, 54,826 for DFG, and 42,572 for JSPS; a paper may be founded by multiple agencies). To control for the topic of research in our comparison, we selected unfunded papers from the same journals and years as the published funded papers. The dotted gray line shows the mean value of *DP* for all the analyzed articles

**Acknowledgments:**
We thank Eamon Duede, Linzhuo Li, Chengjun Wang, and Yang Yang for helpful discussions and comments, Dongbo Li and Jiang Li for sharing Nobel Prize winner data. We thank Clarivate Analytics for supplying the Web of Science data. This work was supported by DARPA's Big Mechanism program grant #14145043, the John Templeton Foundation's grant to the Metaknowledge Network, National Science Foundation grant SBE1158803 and AFOSR grants FA9550-15-1-0162 and FA9550-17-1-0089.